\documentclass[%
 reprint,
superscriptaddress,
 amsmath,
 amssymb,
 aps,
 aas,
 pra,
 longbibliography,
 color,
 tabulary,
 floatfix
]{revtex4-1}

\usepackage[usenames,dvipsnames,svgnames,table]{xcolor}
\usepackage{natbib}
\usepackage{hyperref}
\hypersetup{colorlinks=true,
    linkcolor=blue,
    filecolor=magenta,      
    urlcolor=cyan}
\usepackage{graphicx}
\usepackage{dcolumn}
\usepackage{tabulary}
\usepackage{bm}
\usepackage{hyperref}
\usepackage{siunitx}
\usepackage{longtable}
\usepackage{xcolor}
\usepackage{enumitem}
\newlist{indenteddesc}{description}{1}
\setlist[indenteddesc]{
  leftmargin=0.8in,  
  rightmargin=0in,
  labelindent=0in, 
  labelwidth=0.5in,
  labelsep=0.1in
}

\newcommand{\blue}{\mathrm{blue}}
\newcommand{\red}{\mathrm{red}}

\begin{document}

\title{Astronomical random numbers for quantum foundations experiments}

\author{Calvin Leung}
\email{cleung@g.hmc.edu}
\affiliation{Harvey Mudd College, Claremont, California 91711, USA}
\author{Amy Brown}
\email{afbrown@g.hmc.edu}
\affiliation{Harvey Mudd College, Claremont, California 91711, USA}
\author{Hien Nguyen}%
\email{hien.t.nguyen@jpl.nasa.gov}
\affiliation{NASA Jet Propulsion Laboratory, Pasadena, California 91109, USA}

\author{Andrew S. Friedman}
\email{asf@ucsd.edu}
\affiliation{
University of California, San Diego, La Jolla, California 92093, USA}

\author{David I. Kaiser}
\email{dikaiser@mit.edu}
\affiliation{Massachusetts Institute of Technology, Cambridge, Massachusetts 02139, USA}

\author{Jason Gallicchio}
\email{jason@hmc.edu}
\affiliation{Harvey Mudd College, Claremont, California 91711, USA}

\date{\today}

\begin{abstract}
Photons from distant astronomical sources can be used as a classical source of randomness to improve fundamental tests of quantum nonlocality,
wave-particle duality, and local realism through Bell's inequality and delayed-choice quantum eraser tests inspired by Wheeler's cosmic-scale Mach-Zehnder interferometer gedankenexperiment. Such sources of random numbers may also be useful for information-theoretic applications such as key distribution for quantum cryptography. Building on the design of an ``astronomical random number generator" developed for the recent ``cosmic Bell" experiment \cite{handsteiner2017cosmic}, in this paper
we report on the design and characterization of a device that, with 20-nanosecond latency, outputs a bit based on whether the wavelength of an incoming photon is greater than or less than $\approx \SI{700}{\nano\meter}$. Using the 1-meter telescope at the Jet Propulsion Laboratory (JPL) Table Mountain Observatory, we generated random bits from astronomical photons in both color channels from 50 stars of varying color and magnitude, 
and from 12 quasars with redshifts up to $z = 3.9$. With stars, we achieved bit rates of $\sim \SI{1e6}{\hertz/\meter^2}$, limited by saturation of our single photon detectors, and with quasars of magnitudes between 12.9 and 16, we achieved rates between $\sim 10^{2}$ and $2\times 10^{3} \, {\rm Hz} / {\rm m}^2$.
For bright quasars, the resulting bitstreams exhibit sufficiently low amounts of statistical predictability as quantified by the mutual information. In addition, a sufficiently high fraction of bits generated are of true astronomical origin in order to address both the locality and ``freedom-of-choice" loopholes when used to set the measurement settings in a test of the Bell-CHSH inequality. 
\end{abstract}

\maketitle

\section{Introduction}

Quantum mechanics remains extraordinarily successful empirically, even though many of its central notions depart strongly from those of classical physics. Clever experiments have been designed and conducted over the years to try to test directly such features as quantum nonlocality and wave-particle duality. Many of these tests depend upon a presumed separation between experimenters' choices of specific measurements to perform and features of the physical systems to be measured. Tests of both Bell's inequality and wave-particle duality can therefore make stronger claims about the nature of reality when the measurement bases are determined by events that are separated by significant distances in space and time from the rest of the experiment
\cite{scheidl2010violation,ma2013eraser,GallicchioFriedmanKaiser2014,handsteiner2017cosmic,Wu2016,cao2017bell,yin2017satellite}. 

Bell's inequality~\cite{bell1964einstein} sets a strict limit on how strongly correlated measurement outcomes on pairs of entangled particles can be, if the particles' behavior is described by a local-realist theory. Quantum mechanics does not obey local realism and predicts that for particles in certain states, measurement outcomes can be correlated in excess of Bell's inequality. (In a ``local-realist" theory, no physical influence can travel faster than the speed of light in vacuum, and objects possess complete sets of properties on their own, prior to measurement.) 
Bell's inequality was derived subject to several assumptions, the violation of any of which could enable a local-realist theory to account for correlations that exceed the limit set by Bell's inequality. (For recent discussion of such ``loopholes," see Refs.~\cite{brunner2014,larsson2014,KoflerGiustinaLarsson2016}.) Beginning in 2015, several experimental tests have found clear violations of Bell's inequality while simultaneously closing two of the three most significant loopholes, namely, ``locality" and ``fair sampling" \cite{hensen15,giustina2015significant,shalm2015strong,rosenfeld17}. To close the locality loophole, one must ensure that no information about the measurement setting or outcome at one detector can be communicated (at or below the speed of light) to the second detector before its own measurement has been completed. To close the fair-sampling loophole, one must measure a sufficiently large fraction of the entangled pairs that were produced by the source, to ensure that any correlations that exceed Bell's inequality could not be accounted for due to measurements on some biased sub-ensemble. 

Recent 
work has revived interest in a third major loophole, known as the ``measurement-independence," ``settings-independence," or ``freedom-of-choice" loophole. According to this loophole, local-realist theories that allow for a small but nonzero correlation between the selection of measurement bases and some ``hidden variable" that affects the measurement outcomes are able to mimic the predictions from quantum mechanics, and thereby violate Bell's inequality \cite{scheidl2010violation,GallicchioFriedmanKaiser2014,handsteiner2017cosmic,Wu2016,hall2010local,hall11,barrett11,banik12,putz14,putz16,hall16a,pironio15}. 

A ``cosmic Bell'' experiment was recently conducted that addressed the ``freedom-of-choice" loophole \cite{handsteiner2017cosmic}. A statistically significant violation of Bell's inequality was observed in measurements on pairs of polarization-entangled photons, while measurement bases for each detector were set by real-time astronomical observations of light from Milky Way stars. 
(This experiment also closed the locality loophole, but not fair sampling.) 
The experiment reported in Ref.~\cite{handsteiner2017cosmic} is the first in a series of tests which aim to use the most cosmologically distant sources of randomness available, thus minimizing the plausibility of correlation between the setting choices and any hidden-variable influences that can affect measurement outcomes.

Random bits from cosmologically distant phenomena can also improve tests of wave-particle duality. Wheeler~\cite{wheeler78,wheeler83,miller1984delayed} proposed a ``delayed-choice" experiment in which the paths of an interferometer bent around a distant quasar due to gravitational lensing. By making the choice of whether or not to insert the final beam splitter at the last instant, the photons end up behaving as if they had been particles or waves all along. (For a recent review, see Ref.~\cite{MaKoflerZeilinger16}.) In Section~\ref{sec:eraser}, we will discuss how to feasibly implement an alternative experiment with current technology that retains the same spirit and logical conclusion as Wheeler's original gedankenexperiment.

Beyond such uses in tests of the foundations of quantum mechanics, low-latency astronomical sources of random numbers could be useful in information-theoretic applications as well. For example, such random bits could be instrumental for device-independent quantum-cryptographic key-distribution schemes (as also emphasized in Ref.~\cite{Wu2016}), further solidifying protocols like those described in Refs.~\cite{barrett05,pironio09,pironio10,colbeck12,gallego13,vazirani14,yin2017satellite,winick2017reliable,trushechkin2018reliable,liao2018satellite,lee2017cosmic}. 

In this paper, we describe the design choices and construction of a low-latency astronomical random number generator, building on experience gained in conducting the recent ``cosmic Bell" experiment \cite{handsteiner2017cosmic}. While previous work has successfully generated randomness from astronomical images by reading out the pixels of a CCD camera~\cite{pimbblet2005random}, our unique nanosecond-latency, single-photon instrumentation and our analysis framework make this scheme well-suited for conducting experiments in quantum foundations.
In Section \ref{sec:closing} we formalize and quantify what is required to close the freedom-of-choice loophole in tests of Bell's inequality. This sets a minimum signal-to-noise ratio, which in turn dictates design criteria and choices of astronomical sources. In Section \ref{sec:eraser} we describe how astronomical random number generators may be utilized in realizations of delayed-choice gedankenexperiments, to dramatically isolate the selection of measurements to be performed from the rest of the physical apparatus. In Section \ref{sec:genastrorandom} we compare different ways to turn streams of incoming astronomical photons into an unpredictable binary sequence whose elements were determined at the time of emission at the astronomical source and have not been significantly altered since. After discussing the instrument design in Sections \ref{sec:designconsiderations}-\ref{sec:instrument}, we characterize in Section \ref{sec:calibration} the response of the instrument when observing a number of astronomical targets, including $\approx 50$ bright Milky Way stars selected from the {\sc HIPPARCOS} catalog having different magnitudes, colors, and altitudes. We also describe our observation of 12 quasars with redshifts ranging from $z = 0.1-3.9$. Finally, in Section \ref{sec:quality} we quantify the predictability of the resulting bitstreams, and demonstrate the feasibility of using such quasars in the next round of ``cosmic Bell'' tests. Concluding remarks follow in Section \ref{sec:discussion}.

\section{Closing the Freedom-of-Choice Loophole in Bell Tests\label{sec:closing}}

To address the freedom-of-choice loophole in a cosmic Bell test, the choice of measurement basis on each side of the experiment must be determined by an event at a significant space-time distance from any local influence that could affect the measurement outcomes on the entangled particles \cite{GallicchioFriedmanKaiser2014,handsteiner2017cosmic,yin2017satellite}. As we demonstrate in this section, an average of at least $\approx 79$\% of detector settings on each side must be generated by information that is astronomical in origin, with a higher fraction required in the case of imperfect entanglement visibility. We will label detector settings that are determined by genuinely astronomical events as ``valid,'' and all other detector settings as ``invalid.'' 
We will use this framework to analyze random numbers obtained from both stars and quasars. As we will see in later sections, ``invalid'' setting choices can arise for various reasons, including triggering on local photons (skyglow, light pollution) rather than astronomical photons, detector dark counts, as well as by astronomical photons that produce the ``wrong'' setting due to imperfect optics.

Experimental tests of Bell's inequality typically involve correlations between measurement outcomes $A, B \in \{-1,+1\}$ for particular measurement settings $(a_k, b_\ell)$, with $k,\ell \in \{ 1, 2\}$. Here $a$ and $A$ refer to the measurement setting and outcome at Alice's detector (respectively), and $b$ and $B$ refer to Bob's detector. We follow the notation of Ref.~\cite{handsteiner2017cosmic} and write 
the Clauser-Horne-Shimony-Holt (CHSH) parameter, $S$ \cite{clauser1969proposed}, in the form
\begin{equation}
    S \equiv \vert E_{11} + E_{12} + E_{21} - E_{22} \vert ,
    \label{Sdef}
\end{equation}
where $E_{k\ell} = 2 p (A = B \vert a_k b_\ell) - 1$, and $p ( A = B \vert a_k b_\ell)$ is the probability that Alice and Bob measure the same outcome given the joint settings $(a_k, b_\ell)$. Bell's inequality places a restriction on all local-realist theories. In terms of the quantity $S$, the Bell-CHSH inequality takes the form $S \leq 2$ \cite{clauser1969proposed}.

The value of $S$ that one measures experimentally may be expressed as a linear combination of $S_{\rm valid}$, due to astronomical setting choices, and $S_{\rm invalid}$, due to non-astronomical setting choices. We may write
\begin{equation}
S_{\rm exp} = q S_{\rm valid} + (1-q) S_{\rm invalid} \, ,
\label{eq:s_exp}
\end{equation}
where $q$ is the probability that both setting choices are generated by a given pair of astronomical sources for a given experimental run. 
We conservatively assume that a local-realist theory could exploit the freedom-of-choice loophole to maximize $S_{\rm exp}$ by engineering
each invalid experimental run to yield the mathematical maximum of $S_{\rm invalid} = 4$, while we assume that each valid run
would be limited to $S_{\rm valid} \leq 2$ by the usual Bell-CHSH argument. A ``relaxed'' version of the Bell-CHSH inequality is then $S_{\rm exp} \leq 4 - 2q$.  This makes the statistical significance of any experimental Bell violation highly sensitive to the fraction of valid settings generated. Since quantum mechanics predicts a maximum value 
$S_{\rm QM} = 2\sqrt{2}$~\cite{cirel1980quantum}, and since $S_{\rm exp} \le 4-2q \le S_{\rm QM}$, we conclude that for a cosmic Bell experiment to distinguish between the predictions of quantum mechanics and a local-realist alternative that exploits the freedom-of-choice loophole, we must be able to conduct a sufficiently high fraction $q$ of our experimental runs using valid astronomical photons:
\begin{equation}
q \geq 2-\sqrt{2} \,.
\label{eq:violation}
\end{equation}

In this framework, there are local-realist models in which only one detector's setting choice needs to be influenced or predicted by a hidden-variable mechanism in order to invalidate a given experimental run and produce $S = 4$. We conservatively assume that corrupt settings do not occur simultaneously, allowing the local-realist alternative to maximally exploit each one. If we denote by $q^{(i)}$ the probability that a setting at the $i^{\rm th}$ detector is valid, with $i = ({\rm Alice}, \, {\rm Bob})$, then $(1 - q^{(i)})$ is the probability that the $i^{\rm th}$ detector setting is {\it invalid}. The fraction of valid settings therefore must be at least $q = 1 - (1 - q^{\rm Alice}) - (1 - q^{\rm Bob}) = q^{\rm Alice} + q^{\rm Bob} - 1$. Eq.~(\ref{eq:violation}) may then be written
\begin{equation}
q^{\rm Alice} + q^{\rm Bob} \geq 3 - \sqrt{2}  \> . 
\label{qAqB}
\end{equation} 
For simplicity, if we assume that the experiment is symmetric with $q^{\rm Alice} = q^{\rm Bob} = q^*$, we find that $q^* \geq (3-\sqrt{2})/2 \simeq 79.3\%$. Thus, for a symmetric setup, roughly eight out of ten photons incident on each random number generator need to be of astronomical origin. When choosing a scheme for generating random numbers, it is necessary to keep this ``signal-to-noise'' threshold in mind.

It is also important to consider that it is very difficult in practice to achieve a value of $S$ close to the quantum-mechanical maximum of $2\sqrt{2} \approx 2.83$, due to imperfections in the experimental setup. For example, the first cosmic Bell test obtained values of $S_{\rm exp} = 2.43$ and $S_{\rm exp} = 2.50$~\cite{handsteiner2017cosmic}. Under 
such conditions, $q$ would need
to be correspondingly higher to address the freedom-of-choice loophole. Also, the closer the measurement of $S_{\rm exp}$ is to the validity-modified local-realist bound, the more experimental runs are required to achieve a statistically significant Bell violation. Hence the ``eight-out-of-ten'' rule derived here represents the bare minimum to close the freedom-of-choice loophole for pure entangled states and robust statistics with many experimental runs. In later sections we measure different sources of invalid detections and find quasars that are on both sides of this usefulness bound with our telescope. 

\section{Delayed-Choice Experiments}
\label{sec:eraser}

\begin{figure}
    \centering
    \includegraphics[width = 3.5 in]{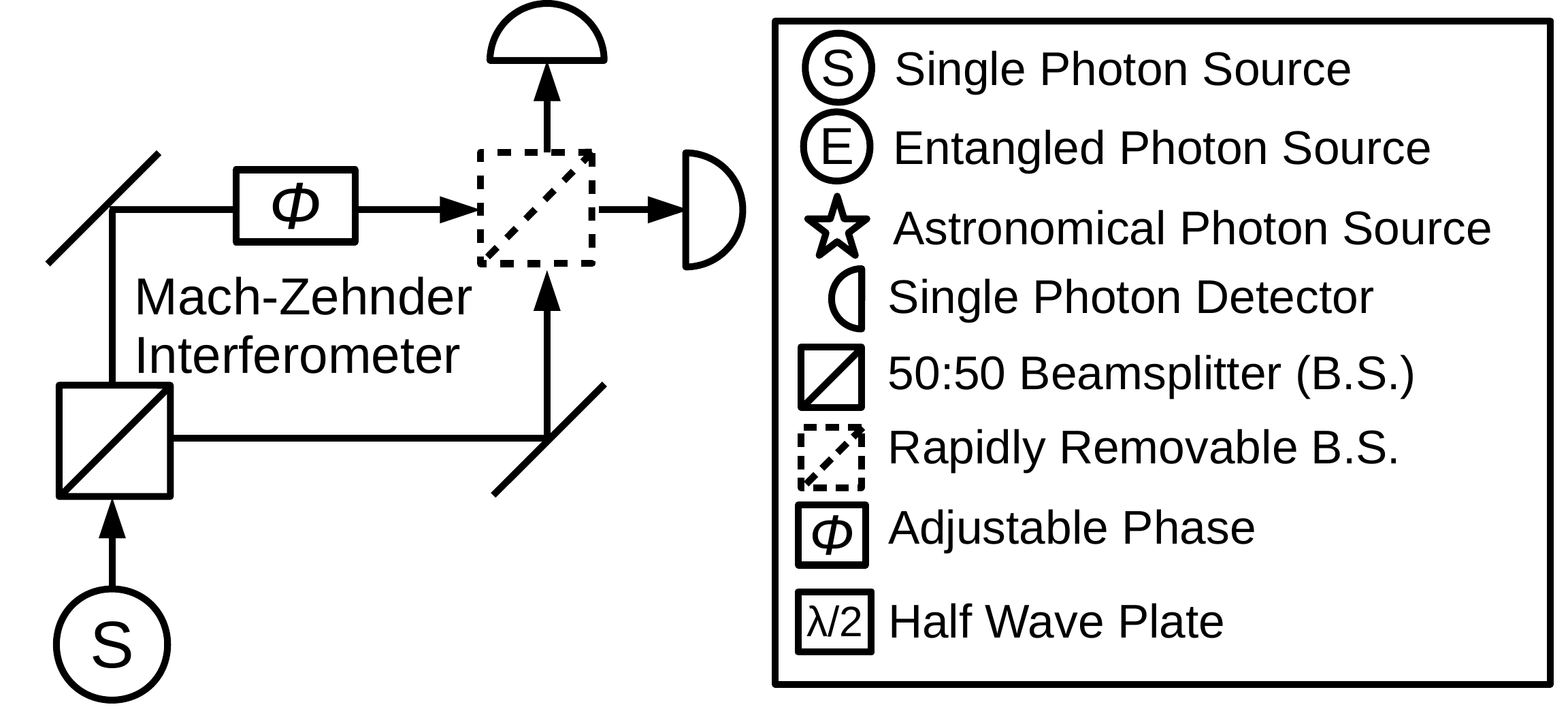}
    \caption{Wheeler's original delayed-choice proposal on a tabletop where the second beamsplitter can be rapidly inserted/removed after a single photon from $S$ passes the first beamsplitter. The legend here applies for Figs.~\ref{fig:wheeler-delay-choice}-\ref{fig:eraser}.}
    \label{fig:wheeler-delay-choice}
\end{figure}

Another application of an astronomical random number generator is to use it in an experiment to test wave-particle duality. The concept of testing wave-particle duality with a Mach-Zehnder interferometer was first proposed by John Archibald Wheeler~\cite{wheeler78,wheeler83} and has been realized in several laboratory-scale experiments using single photons and single atoms~\cite{baldzuhn1989wave,jacques2008delayed,manning2015wheeler}. In such an experiment, each photon that enters the first beamsplitter exhibits self-interference if the second beamsplitter is present, and the pattern of single-photon detections observed after aggregating many trials is in correspondence with a classical wave picture. However if the final beamsplitter is absent, the light from each path would not recombine, and single photons would appear at one output or the other, revealing which path was taken. In Wheeler's original proposal~\cite{wheeler78}, the experimenter would be able to choose whether to insert or remove the second beamsplitter \textit{after} the photon had entered the interferometer. Such a scenario was dubbed a ``delayed-choice'' experiment because the photon's trajectory---one path, the other, or both---was determined after it passed the first beamsplitter. If one rejects wave-particle duality, the logical conclusion is that either the choice of removing the final beamsplitter in the final moments of the light's journey somehow retrocausally affected the light's trajectory, or that the experimenter's choice of removing the final beamsplitter was predictable by the light before it embarked on its journey. (See also Ref.~\cite{MaKoflerZeilinger16}.) See Fig. \ref{fig:wheeler-delay-choice}.

\begin{figure}
    \centering
    \includegraphics[width = 3.5 in]{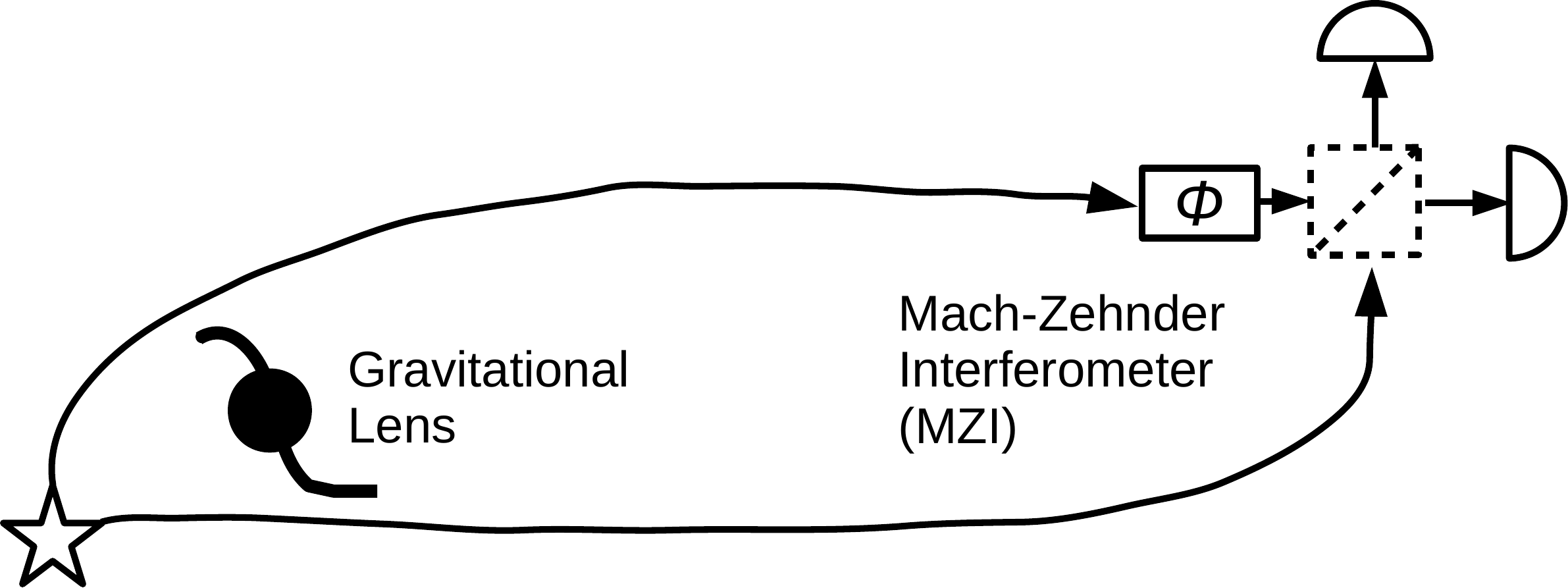}
    \caption{Wheeler's cosmic delayed-choice proposal: using a gravitational lens as an interferometer with a quasar photon taking one or both paths.}
    \label{fig:wheeler-gravitational-lens}
\end{figure}

Wheeler next proposed~\cite{miller1984delayed} a cosmological version of this test, with the source of interfering photons being a cosmologically distant quasar and the first beamsplitter being an intervening gravitational lens that produces at least two images of the quasar on Earth. If the two images are recombined at a final laboratory beamsplitter, the quasar photons would exhibit interference between distinct paths of cosmological scale.  If the final beamsplitter were removed, the photons would not exhibit interference and one could presumably identify unique trajectories for such photons from emission at the quasar to detection on Earth. If one insists on rejecting wave/particle duality in this case, it would appear as if the experimenter's choice on Earth had determined whether the photon took one path or both, billions of years ago. See Fig. \ref{fig:wheeler-gravitational-lens}.

\begin{figure}
    \centering
    \includegraphics[width = 3.5 in]{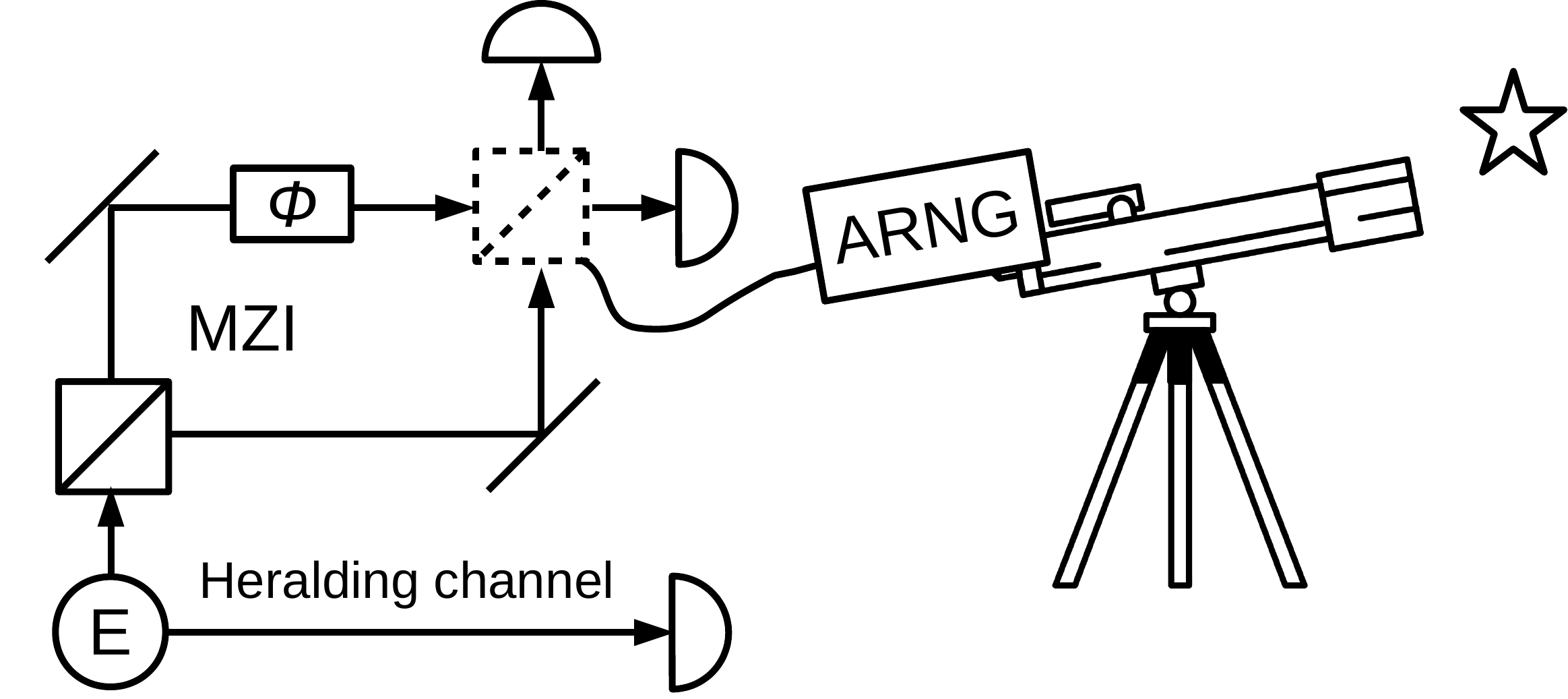}
    \caption{One way to perform a delayed-choice experiment that keeps the spirit of Wheeler's cosmic proposal, using an astronomical random number generator to determine whether to insert/remove the second beamsplitter in the Mach-Zehnder interferometer. Like in Wheeler's original proposal, the spacetime location where the choice is made is separated from the interferometer's first beamsplitter by astronomical distances. However, unlike in Wheeler's gravitational lens proposal, the astronomical photon does not go through the interferometer nor does it exhibit any wave-like properties. It is instead used to generate a classical random number that determines whether to insert or remove the beamsplitter while a locally-generated single photon is in flight. 
    It is helpful to use a photon pair source $E$ to generate single photons and to post-select coincidence events.}
    \label{fig:wheeler-heralded}
\end{figure}

The feasibility of realizing Wheeler's quasar experiment has been explored~\cite{doyle2008quantum}. The central difficulty is maintaining the quantum coherence of the light traveling over cosmological distances.  Rather than try to interfere astronomical photons with a gravitational lens, we can realize a related experiment that leads to the same logical conclusion. Instead of testing the wave-particle duality of an astronomical photon, we may use a standard tabletop Mach-Zehnder interferometer, and use astronomical setting choices to determine whether to insert or remove the beamsplitter after a laboratory-produced photon has entered the interferometer. In such a setup, the choice of which measurement to perform would be made in a causally disconnected way from the particulars of the behavior of the photon in the interferometer, billions of years before the interferometer photon had even been created. See Fig.~\ref{fig:wheeler-heralded}.

In this experiment as well as Wheeler's original gedankenexperiment, a cosmologically long time interval is realized between when a photon enters the first beamsplitter, and when the presence/absence of the second beamsplitter is determined. In Wheeler's experiment, the photon enters the gravitational lens and the second beamsplitter's presence is determined billions of years later by experimenters on Earth. In our proposed experiment, a quasar photon emitted billions of years ago determines the state of the second beamsplitter, while laboratory-generated single photons are sent into a tabletop interferometer.  Separating the choice of inserting the beamsplitter from both the creation of the photon and its journey makes alternate explanations of wave-particle duality implausible.

\begin{figure}
    \centering
    \includegraphics[width = 3.375in]{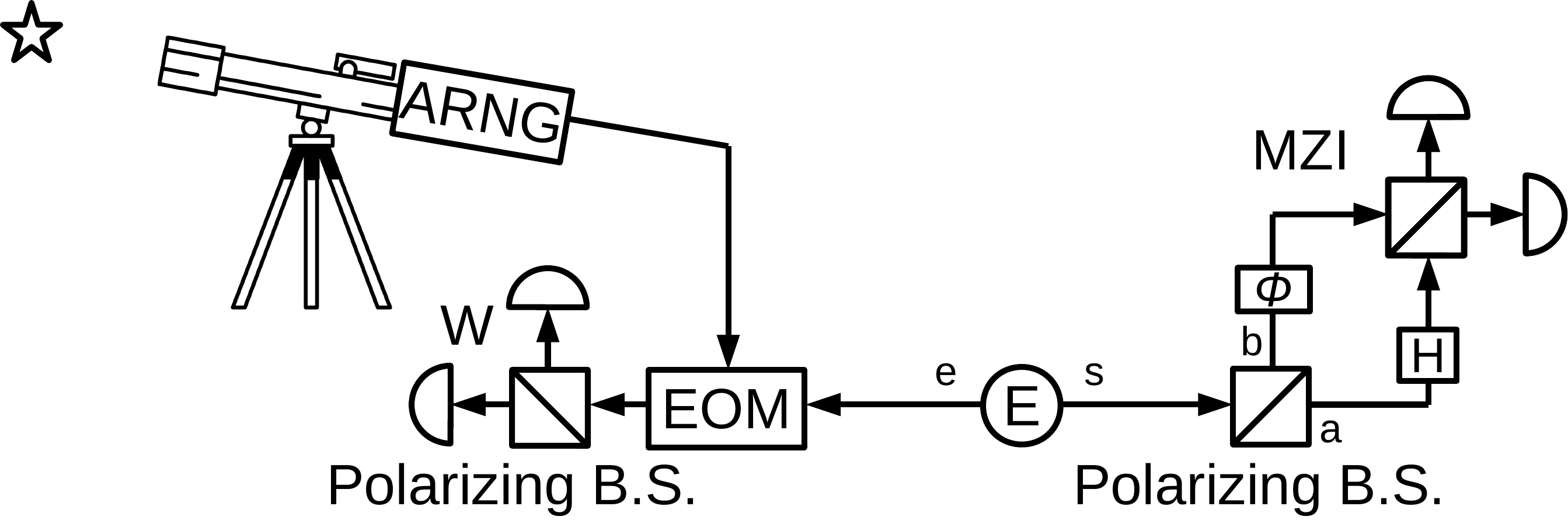}
    \caption{A proposed ``cosmic eraser'' experiment. A two-photon entangled state is produced at $E$, sending one entangled partner (the ``environment" photon) towards $W$ and the other (the ``signal" photon) toward a Mach-Zehnder interferometer (MZI). An astronomical random number generator (ARNG) activates an electro-optical modulator (EOM) in order to rapidly set the measurement basis for the environment photon at $W$, potentially revealing which-path information about the signal photon. The signal photon at the MZI acts as a particle or a wave accordingly, even though the decision point of whether to reveal which-path information is made potentially billions of years before the experiment has been run, out of the past light cone of the interferometer.}
    \label{fig:eraser}
\end{figure}

In addition to such delayed-choice experiments, a related line of experiments probe so-called ``quantum erasure'' \cite{MaKoflerZeilinger16}, which likewise draw inspiration from Wheeler's original proposal (See also \cite{endnotetext,fankhauser2017taming}). In modern delayed-choice quantum-eraser experiments~\cite{ma2013eraser}, wave-particle duality is tested by interfering one entangled partner (the ``signal" photon) of a two-photon entangled state in a Mach-Zehnder interferometer. Rather than removing the beamsplitter in the Mach-Zehnder interferometer, a measurement of the other entangled partner (the ``environment" photon) is made outside the light cone of the signal photon to erase which-path information. This can be done at the same time or after the signal photon propagates through the interferometer~\cite{ma2013eraser,MaKoflerZeilinger16}. Here again, we can realize Wheeler's original ambition to manifest the features of quantum mechanics on cosmic scales in a ``cosmic eraser'' experiment. In our proposed test, light from an astronomical source would determine whether which-way information is erased. See Fig.~\ref{fig:eraser}.

In the framework of quantum mechanics, these quantum eraser experiments begin with a polarization-entangled state of ``signal" and ``environment" photons. Following the discussion in Ref.~\cite{ma2013eraser}, we may write such a state as
\begin{equation}
\vert \psi \rangle = \frac{1}{\sqrt{2}} (\vert H \rangle_s \vert V \rangle_e + \vert V \rangle_s \vert H \rangle_e) .
\end{equation}
When the signal photon enters the interferometer, the polarizing beamsplitter maps the polarization information of the signal photon onto which path it takes through the interferometer, with horizontally polarized photons taking path $a$ and vertically polarized photons taking path $b$. A half-wave plate rotates path $a$'s horizontal polarization into vertical polarization, erasing which-way information encoded in the polarization of this photon:  $|H\rangle \to |a,V\rangle = |a\rangle$ and $|V\rangle \to |b,V\rangle = |b\rangle$. If we assume the $b$ path picks up an adjustable phase $\phi$, the state afterward may be written as
\begin{align}
\vert \psi &\rangle \rightarrow \frac{1}{\sqrt{2}} ( \vert a\rangle_s \vert V \rangle_e + e^{i\phi} \vert b \rangle_s \vert H \rangle_e )\\
&= \frac{1}{ 2} \left[ ( \vert a \rangle_s + i e^{i\phi} \vert b \rangle_s ) \vert L \rangle_e + (\vert a \rangle_s - i e^{i\phi} \vert b \rangle_s ) \vert R \rangle_e \right] .
\end{align}

After the final 50/50 beamsplitter in the interferometer, the two signal paths will recombine. The signal's which-way information is still potentially available in the polarization of the environment photon. If the environment photon is measured in the $| H\rangle_e, | V\rangle_e$ basis, which-path information about the signal photon is nonlocally revealed, and no phase-dependent interference is observed. We can see this in the joint probability of any pair of signal and 
environment detectors firing simultaneously: the probability that both upper detectors register a coincidence when measuring in the
$| H\rangle_e, | V\rangle_e$ basis is

\begin{equation} 
P_{HV} = \dfrac{1}{2}|\langle V|_e\langle a|_s |\psi \rangle + \langle V|_e \langle b|_s |\psi \rangle|^2 = \dfrac{1}{4}
\end{equation}
and no interference fringes are observed in the coincidence probability. On the other hand, if the electro-optic modulator (EOM) rotates the environment photon such that incoming $|R \rangle_e$ photons enter the upper detector and incoming $|L \rangle_e$ photons enter the lower detector, information about the signal photon's path is lost. Then the coincidence probability is given by 
\begin{equation} 
P_{LR} = \dfrac{1}{2}|\langle R|_e\langle a|_s |\psi \rangle + \langle R|_e \langle b|_s |\psi \rangle|^2 = \dfrac{1}{4}(1 + \sin \phi )
\end{equation}
and interference fringes are observed in the coincidence probabilites. 
We emphasize that for both linear and circular basis choices, the signal photon enters each detector with equal probability, so as with any entangled state, information cannot be sent simply by nonlocally choosing a measurement basis. Interference fringes or the lack thereof can only be seen when one sorts the signal photon's detections into categories based on the basis choice and measurement result of the environment photon. As in tests of Bell's inequality, any apparent nonlocality is only nonlocality of correlations.

Any local explanation of the nonlocal correlations in this experiment would rely on being able to predict whether the measurement of the environment photon erases or reveals which-path information of the signal photon, dictating the wave-like or particle-like behavior of the signal photon. Setting the environment photon's measurement basis with a single astronomical random number generator can be used to dramatically constrain the potential origins of this predictability.

\section{Generating Astronomical Randomness}
\label{sec:genastrorandom}

We consider 
two potential schemes
for extracting bits of information from astronomical photons to use as sources of randomness for use in experiments like those described in Sections \ref{sec:closing}-\ref{sec:eraser}. In general, it is important that the information extracted be set at the time of the astronomical photon's emission, rather than at the time of detection or any intervening time during the photon's propagation. We deem the setting corrupt if this condition is not met, and we evaluate two methods with particular emphasis on the mechanisms by which corruption may occur.

\subsection{Time of Arrival}
The first method is to use the time-of-arrival of the astronomical photons to generate bits \cite{GallicchioFriedmanKaiser2014,Wu2016}. We can choose to map time tags to bits based on whether some pre-specified decimal place of the timestamp is even or odd. For example, a $0$ could correspond to the case of a photon arriving on an even nanosecond, and a $1$ for arrival on an odd nanosecond. The main advantage of this scheme is its simplicity: since timestamps need to be recorded to close the locality loophole, there is no need for additional hardware to generate random settings. In addition, it will always be possible to ensure a near-50/50 split between the two possible setting choices at each side of the experiment regardless of the source of astronomical randomness. Indeed, our time tags, when mapped to random bits by their timestamp, pass every test of randomness in the NIST Statistical Test Suite for which we had sufficient bits to run them \cite{rukhin2001statistical}.

The primary disadvantage of this scheme is that it is very difficult to quantify galactic and terrestrial influences on the recorded timestamp of the photon's arrival. It is necessary that we be able to quantify the fraction of photons that are corrupt, as discussed in Section~\ref{sec:closing}. In the remainder of this section, we consider the constraints on which decimal place in the detection timestamp should be used to generate random bits.

It is tempting to condition setting choices on the even/oddness of a sub-nanosecond decimal place, making use of deterministic chaos and apparent randomness. However, the timestamp of a given photon's arrival at this level of precision is sensitive to corruption from myriad local influences which are difficult (perhaps impossible) to quantify, such as effects in the interstellar medium, time-dependent atmospheric turbulence, and timing jitter in the detectors or time-tagging unit, which may affect the even-odd classification of nanosecond timestamps. The atmosphere has an index of refraction $n \approx 1 + 2.9\times 10^{-4}$, which in a $\SI{10}{\km}$-thick atmosphere corresponds to the photons arriving $\sim \SI{10}{\ns}$ later than they would if traveling in a vacuum \cite{owens1967optical}. Thus, relying upon any decimal place less significant than the tens-of-nanoseconds place to generate a bit admits the possibility of the atmosphere introducing some subtle delay and corrupting the generated bits. 

Choosing a setting by looking at the even/oddness of microsecond timestamps, on the other hand, makes it difficult to close the locality loophole in tests of Bell's inequality. To close the locality loophole, a random bit must be generated on each side of the experiment within a single timing window, whose duration is set by the distance between the source of entangled particles and the closer of the two measurement stations ($\approx \SI{3}{\us}$ in the first cosmic Bell experiment~\cite{handsteiner2017cosmic}).  The coincidence rate between the two RNGs is proportional to the bit generation rate on each side, increasing the number of Bell runs achievable within a certain experiment runtime. However, if the bit generation rate increased, the bits lose their apparent randomness: generating bits at any rate faster than $\SI{1}{\us}^{-1}$
would simply yield strings of consecutive 0's and 1's. This creates a difficult scenario where the experimenter can only increase the rate of successful runs by sacrificing the statistical unpredictability of the random bits, in a scenario where it is already desirable to maximize the rate of successful runs due to practical constraints on observatory telescope time.

In addition, for rates that are slow compared to the causal validity time, the remote setting choice on each side of the experiment is a deterministic function of time. Using even/odd timestamps to determine the setting choice admits the possibility that a local hidden variable theory, acting at the entanglement source, emits photon pairs to coincide with a particular setting choice. For these reasons, using the timestamp of astronomical photons' arrivals does not appear to be an optimal method for generating unpredictable numbers of astronomical origin.

\subsection{Colors}
\label{sec:colors}

An alternate approach, developed for use in the recent cosmic Bell test \cite{handsteiner2017cosmic}, is to classify astronomical photons by designating a central wavelength $\lambda'$ and mapping all detections with $\lambda < \lambda'$ to 0 and detections with $\lambda > \lambda'$ to 1 using dichroic beamsplitters with appropriately chosen spectral responses. The advantage of the wavelength scheme is that possible terrestrial influences on photons as a function of wavelength are well-studied and characterized by empirical studies of astronomical spectra, as well as studies of absorption and scattering in the atmosphere. In contrast to effects which alter arrival times, the effects of the atmosphere on the distribution of photon wavelengths varies over the course of minutes or hours, as astronomical sources get exposed to a slowly-varying airmass over the course of a night-long Bell test. The airmass, and therefore the atmosphere's corrupting influence on incoming astronomical photons, can be readily quantified as a function of time.  

One important advantage of using astronomical photons' color stems from the fact that in an optically linear medium, there does not exist any known physical process that could absorb and re-radiate a given photon at a different wavelength {\it along our line of sight}, without violating the local conservation of energy and momentum \cite{handsteiner2017cosmic}. While photons could scatter off particles in the intergalactic media (IGM), interstellar media (ISM), or Earth's atmosphere, a straightforward calculation of the column densities for each medium indicates that among these, the number of scatterers per square meter is highest in the Earth's atmosphere by more than two orders of magnitude compared to the ISM in the Milky Way, and several orders of magnitude greater than in the IGM~\cite{madau2000intergalactic}. Hence, treating the IGM and ISM as transparent media for photons of optical frequencies from distant quasars is a reasonable approximation.

For photons of genuinely cosmic origin, certain well-understood physical processes do alter the wavelength of a given photon between emission and detection, such as cosmological redshift due to Hubble expansion. Such effects, however, should not be an impediment to using astronomical photons' color to test local-realist alternatives to quantum mechanics.

The effects of cosmological redshift are independent of a photon's wavelength at emission, and hence treat all photons from a given astronomical source in a comparable way \cite{Peeblesbook,Weinbergbook}. Gravitational lensing effects are also 
independent of a photon's wavelength at emission \cite{blandford92}, though lensing accompanied by strong plasma effects can yield wavelength-dependent shifts \cite{rogers15}. Even in the latter case, however, any hidden-variable mechanism that might aim to exploit gravitational lensing to adjust the detected wavelengths of astronomical photons on a photon-by-photon basis would presumably need to be able to manipulate enormous objects (such as neutron stars) or their associated magnetic fields (with field strengths $B > 10^8$ Gauss) with nanosecond accuracy, which would require the injection or removal of genuinely astronomical amounts of energy. Thus, whereas some of the original hidden-variable models were designed to account for (and hence be able to affect) particles' trajectories \cite{Bellbook,bush15} --- including, thereby, their arrival times at a detector --- any hidden-variable mechanism that might aim to change the color of astronomical photons on a photon-by-photon basis would require significant changes to the local energy and momentum of the system.

The chief disadvantage of using photons' color in an astronomical random number generator is that the fluxes of ``red'' ($\lambda > \lambda'$) and ``blue'' ($\lambda < \lambda'$) photons will almost never be in equal proportion, and hence will yield an overall red-blue statistical imbalance. Such an imbalance in itself need not be a problem: one may conduct Bell tests with an imbalance in the frequency with which various detector-setting combinations are selected 
\cite{KoflerGiustinaLarsson2016,handsteiner2017cosmic}. However, a large red-blue imbalance does affect the duration of an experiment --- whose duration is intrinsically limited by the length of the night --- because collecting robust statistics for each of the four joint setting choices $(a_k, b_\ell)$ would prolong the experiment.

A second disadvantage comes from imperfect alignment. If the detectors for different colors are sensitive to different locations on the sky, atmospheric turbulence can affect the paths of photons and the relative detection rates. We see evidence of this effect at the sub-percent-level in 
the measurements described in Sections \ref{sec:calibration}: the probability of the next photon being the same color as the previous few photons slightly exceeds what is expected from an overall red-blue imbalance. We quantify this effect in terms of mutual information in Section \ref{sec:quality}. This effect could have been mitigated through better alignment since our aperture was smaller than the active areas of our detectors, but the sensitivity profiles of our detectors' active areas would have to be identical to eliminate it entirely.

We devote the remainder of this paper to the photon-color scheme, given its advantages over the timestamp scheme. We point out that any time-tagging hardware that outputs bits based on color can also output bits based on timing. 


\section{Design Considerations}
\label{sec:designconsiderations}

As became clear during the preparation and conduct of the recent cosmic Bell experiment \cite{handsteiner2017cosmic}, in
designing an instrument that uses photon colors to generate randomness, it is necessary to begin with a model of how settings become corrupted by local influences, and make design choices to minimize this. In this section we build on the discussion in Ref.~\cite{handsteiner2017cosmic} to characterize valid and invalid settings choices.

One obvious source of potential terrestrial corruption is from background noise, due to thermal fluctuations in the detector (or ``dark counts''), as well as background light from the atmosphere (or ``skyglow''). We designate the sum of these two rates as $n_j^{(i)}$, where $j$ labels the two detector arms (red and blue) and $i$ labels the two random number generators (Alice and Bob) in a test of Bell's inequalities. If we measure a count rate of $r^{(i)}_{j}$ when pointing at an astronomical source, then the probability of obtaining a noise count is simply $n^{(i)}_{j} / r^{(i)}_{j}$. In selecting optics, it is important to select single-photon detectors which have low dark count rates and a small field of view on the sky in order to minimize this probability. 

A second source of terrestrial corruption is misclassification of photon colors. A typical way to sort photons by color is to use a dichroic beamsplitter. However, due to imperfections in the dichroic beamsplitter's spectrum, there is a nonzero probability that a photon in the ``red'' wavelength range is transmitted towards the arm designated for ``blue'' photons and vice versa. We need to select dichroic beamsplitters with high extinction ratios and steep transitions such that crosstalk is minimized. 

To quantify the contribution from imperfect dichroic mirrors, we define $j'$ to be the color opposite to $j$, that is, red if $j$ refers to blue and vice versa. Depending on the source spectrum, some fraction $f^{(i)}_{j' \to j}$ of photons end up in the $j^{\rm th}$ arm, despite being of the $j'^{\rm th}$ color. If $s^{(i)}_{j}$ astronomical photons per second of color $j$ are intended for the $i^{\rm th}$ detector, photons leak into the $j'^{\rm{th}}$ arm at a rate of $f_{j \to j'} s^{(i)}_{j}$. Knowing $r^{(i)}_{j}$, $n^{(i)}_j$, as well as the mixture rates $f_{j' \to j}, f_{j \to j'}$ allows us to ``unmix'' the observed count rates $r_{j}$ to back out the true fluxes $s^{(i)}_{j}$. We will discuss the computation of $f^{(i)}_{j' \to j}$ for our instrument in a later section.

In summary, the rate that the $j^{\rm th}$ detector arm in the $i^{\rm th}$ detector yields a corrupt setting is at most the sum of the noise rate, $n^{(i)}_j$, and the rate of misclassifications from the $j'^{\rm th}$ arm, $f_{j' \to j} s_{j'}^{(i)}$. Since the total observed count rate is $r_{j}^{(i)}$, the probability of obtaining an incorrect setting is
\begin{equation}
p^{(i)}_{j} = \frac{n^{(i)}_{j} }{ r^{(i)}_{j} } + \frac{ s^{(i)}_{j'} f_{j' \to j} }{ r^{(i)}_{j} } \,.
\end{equation}
The overall probability of corruption for a bit is conservatively estimated by maximizing over its red and blue detector arms. Since the overall probability of corruption is not necessarily the same for Alice and Bob, we denote this invalid-bit probability $p^{(i)}$, where
\begin{equation}
p^{(i)} = \max ( p^{(i)}_{\textrm{red}}, p^{(i)}_{\textrm{blue}} ) = 1 - q^{(i)}\,,
\end{equation}
where the average of the two valid-bit probabilities $q^{(i)}$ needs to be at least 79.3\%, as discussed in Section \ref{sec:closing}. Note that the $j$ index labels individual detector arms, whereas the $i$ index labels different  observers' detectors after maximizing over each detector's arms.

To minimize an individual detector arm's corruption probability $p_{j}$, it suffices to minimize the quantities $n_{j}$ by minimizing the dark count and skyglow rates, and to choose high-quality dichroic beamsplitters to minimize $f_{j' \to j}$. The total count rate, $r_j$, is maximized when the atmosphere is most transparent: thus, we will designate our red and blue observing bands to roughly coincide with the near-infrared ($\SI{700}{\nm}-\SI{1150}{\nm}$) and optical ($\SI{350}{\nm}-\SI{700}{\nm}$) respectively
\cite{GallicchioFriedmanKaiser2014,handsteiner2017cosmic}.

Several other design considerations are equally important. The instrument must be able to point to dim and distant target objects, which are typically high-redshift quasars. The dimness of even the brightest high-redshift quasars in optical and near-infrared (NIR) wavelengths not only makes it difficult to establish the high signal-to-noise ratio required, but also makes tracking objects nontrivial over the duration of a Bell test, which can last for hours. At the same time, the instrument must generate settings at a sufficiently high rate to perform the experiment. Each run of a Bell inequality test only closes the locality and freedom-of-choice loopholes if valid settings from quasars arrive on both sides within a time window whose duration is set by the light-travel time between Alice and Bob. Thus having a high collection efficiency of the quasar light is doubly important.

\section{Instrument}
\label{sec:instrument}

\begin{figure}
    \centering
    \includegraphics[width = 3.375in]{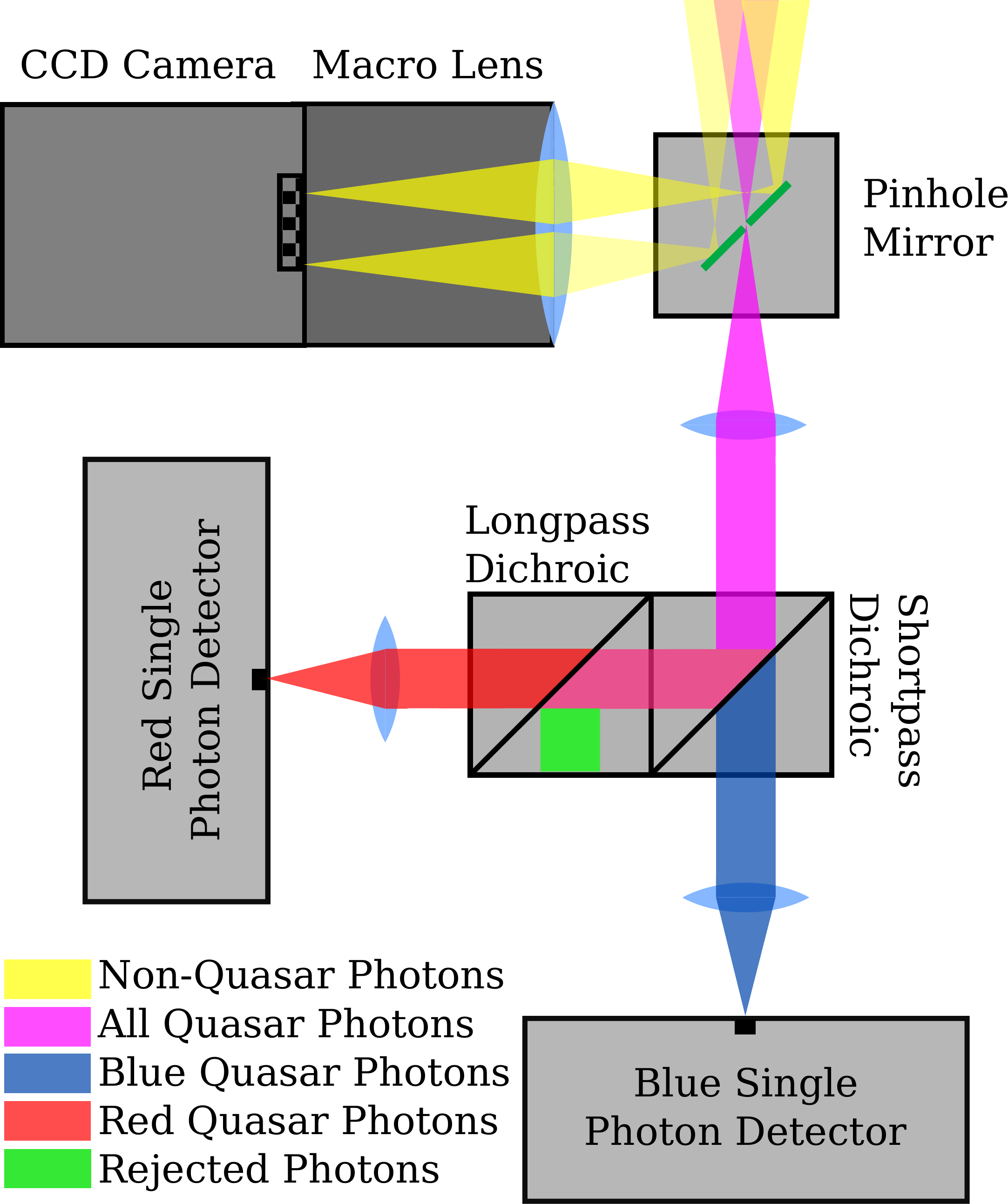}
    \caption{This figure shows the intended optical paths of our astronomical random number generator (not to scale). Astronomical light from multiple objects in the field of view of the telescope enters at the top right of the schematic. This light is brought to a focus by the telescope onto the plane of the pinhole mirror. Most of the light is reflected by the mirror (yellow) and refocused onto a CCD. However, light from an object of interest (purple) passes through the pinhole, and is then collimated and sorted by color via a system of one shortpass and one longpass dichroic beamsplitter. These beams (red and blue) are refocused onto the active area of our two avalanche photodiodes for detection and timestamping. The placement of the dichroics is similar to the fiber-coupled scheme used in Ref.~\cite{handsteiner2017cosmic}.} 
    \label{fig:crngSchematic}
\end{figure}

\begin{figure}
    \centering
    \includegraphics[width = 3.375in]{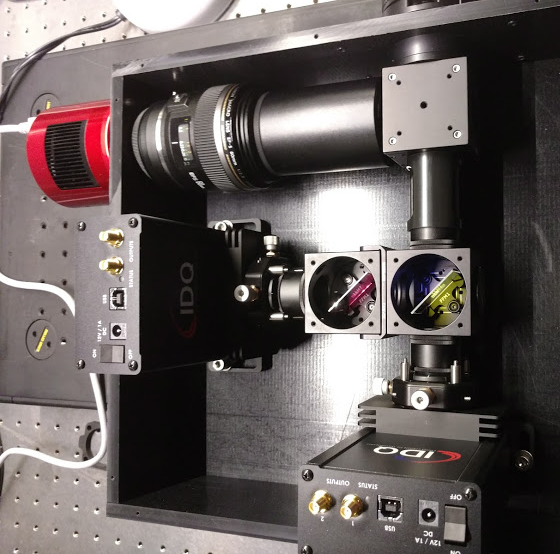}
    \caption{Photo of our astronomical
    random number generator in the laboratory with the lid off and dichroic beamsplitters exposed. 
    }
    \label{fig:crngPhoto}
\end{figure}

Our astronomical random number generator incorporates several design features that were developed in the course of preparing for and conducting the recent cosmic Bell experiment \cite{handsteiner2017cosmic}.
A schematic of our new instrument, constructed at the Harvey Mudd College Department of Physics, is shown in Fig.~\ref{fig:crngSchematic} and a photo in Fig.~\ref{fig:crngPhoto}. It is housed in a box made of black Delrin plastic of dimensions $30\times 30\times 10$ centimeters and weighs $\SI{5.5}{\kg}$, most of which is the weight of two single-photon detectors and the astronomical pointing camera. 
The instrument was mounted at the focus of a 1-meter aperture, 15-meter focal-length telescope at the NASA Jet Propulsion Laboratory's Table Mountain Observatory. The light from the telescope is coupled directly into our instrument's aperture without using optical fibers to reduce coupling losses.
\subsection{Optics}
\begin{figure}
    \centering
    \includegraphics[width = 3.375in]{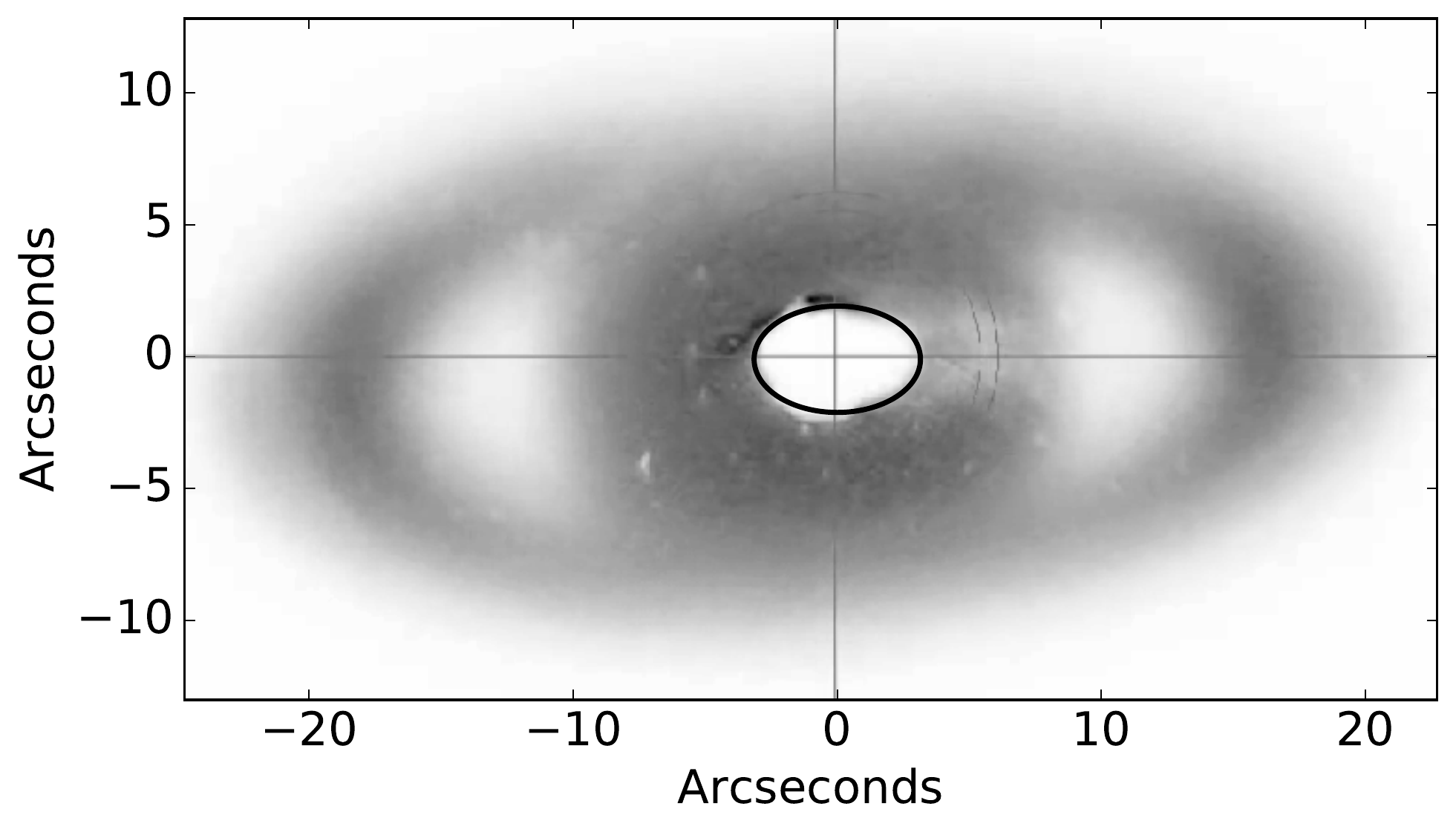}
    \caption{Using the date of observation (3 July 2016) and the coordinates of Table Mountain Observatory, it is possible to compute the angular diameter of Saturn. This enables us to estimate the size of the pinhole as an ellipse with semimajor axes of 2.01 and 3.15 arcseconds. The horizontal and vertical lines running through the pinhole are crosshairs to guide the eye. The field of view calculated via Saturn is consistent with the field of view computed using telescope and camera parameters.}
    \label{fig:saturn}
\end{figure}

\begin{figure}
    \centering
    \includegraphics[width = 3.375in]{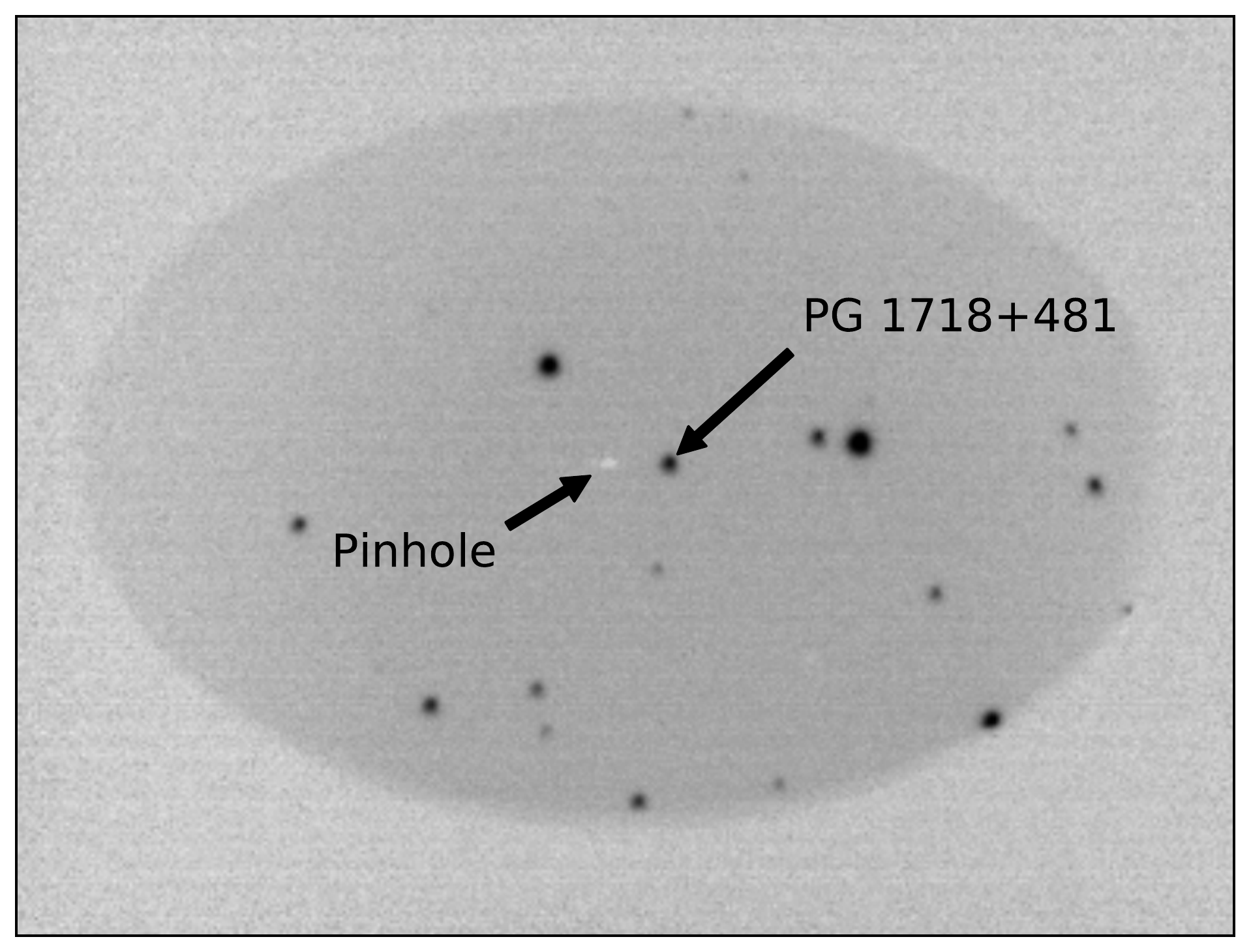}
    \caption{Dim objects such as the quasar PG 1718 + 481 (shown here) were identified by comparing the local field to astronomical catalogs. Dark counts were typically recorded by keeping the object a few spot-sizes away from the pinhole, for example, as the telescope is positioned here.}
    \label{fig:quasarField}
\end{figure}

The telescope light is focused onto a 200 $\mu$m pinhole on a Lenox Laser 45$^\circ$ pinhole mirror. The size of this pinhole was chosen to minimize skyglow background (and therefore the predictability due to skyglow) by matching the 2-3\,arcsecond astronomical seeing at the Table Mountain site. The pinhole diameter corresponds to 2.75\,arcseconds on our 15\,m focal-length telescope. The incoming light that does not pass through the pinhole is reflected by the mirror and re-imaged through a Canon EF-S 60mm F2.8 macro lens onto a ZWO ASI 1600MM cooled 4/3" CMOS camera, which aids in finding and positioning the source into the pinhole. Real-time monitoring of this camera was used to guide the telescope in some observations and to capture long exposures as in Fig.~\ref{fig:saturn} and Fig.~\ref{fig:quasarField}.


The light from the object of interest that passes through the pinhole gets collimated by a 25\,mm diameter, 50\,mm focal-length achromatic lens (Edmund 49-356-INK). This collimated light gets split by a system of two dichroic beamsplitters, with shorter-wavelength light (denoted ``blue'') being transmitted and longer-wavelength light being reflected. The beams are focused onto one IDQ ID120 Silicon Avalanche Photodiode detector through a 25\,mm diameter, 35\,mm focal-length achromatic lens (Edmund 49-353-INK) mounted on a two-axis translation stage attached to the detector. The image of the pinhole is reduced to 140\,$\mu$m in diameter, which is well within the ID120's 500\,$\mu$m active area, making for stable alignment and minimal concern about aberrations and diffraction. The efficiency of the whole system|from the top of the atmosphere to an electronic pulse|is on the order of 30\%, dominated by loss from the detectors and Rayleigh scattering in the atmosphere.

\subsection{Detectors and Time Tagging}
The ID120 Silicon Avalanche Photodiode Detectors (APDs) have up to 80\% quantum efficiency between 350 and 1000\,nm and a low ($<100$\,Hz) specified dark count rate. These have an artificially extended deadtime of $\SI{420}{\ns}$ to prevent afterpulsing. They have a photon-to-electrical-pulse latency of up to 20 ns. The detectors' active area was cooled to $\SI{-40}{\celsius}$ and achieved a measured dark count rate of $\approx \SI{40}{\hertz}$. 
Signals from the APDs are recorded by an IDQ ID801 Time to Digital Converter (TDC). The relative precision of time-tags is limited by the $\SI{80.955}{\ps}$ clock rate of the TDC, and by the \SI{300}{\ps} timing jitter on the APD.  As a timing reference, we also record a stabilized 1-pulse-per-second signal from a Spectrum Instruments TM-4 GPS unit. (Absolute time can also be recorded using this GPS unit's IRIG-B output.) The GPS timing solution from the satellites is compensated for the length of its transponder cable, which corresponds to a delay of $\SI{77}{\ns}$.

\begin{figure}
    \centering
    \includegraphics[width = 3.375in]{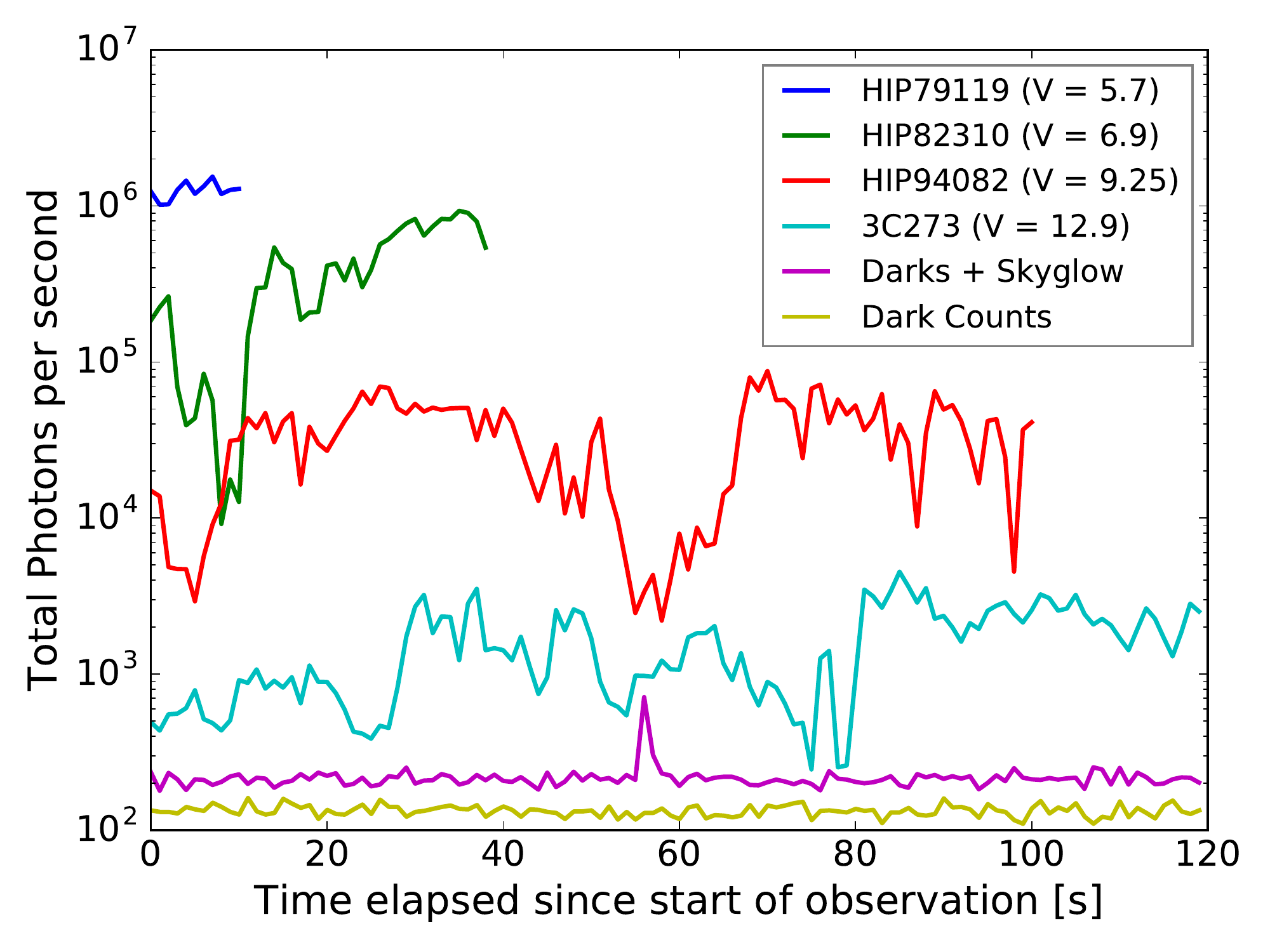}
    \caption{The total count rate over time for various sources fluctuates dramatically due to 2-3 arcseconds of seeing and telescope pointing, which are on the order of our pinhole size. The legend entries appear in the same vertical ordering as on the plot. The small spike in the ``Darks + Skyglow'' trace is likely from an object such as a plane or satellite that briefly passed through our field of view, or headlights from a car.}
    \label{fig:countsOverTime}
\end{figure}

\subsection{Dichroic Beamsplitters}
\label{sec:spectra}

Building on the analysis in Ref.~\cite{handsteiner2017cosmic}, we formulate a model of the instrument's spectral response in each arm to characterize its ability to distinguish red from blue photons. The aim of this section is to compute the $f_{j\to j'}$ parameters for our instrument, defined as the probability that photons of type $j$ are detected as 
photons of type $j'$. As described in Section \ref{sec:designconsiderations}, such misclassified photons contribute to ``invalid" detector-setting choices in the same way that noise does. 

\begin{figure}
    \centering
    \includegraphics[width = 3.375in]{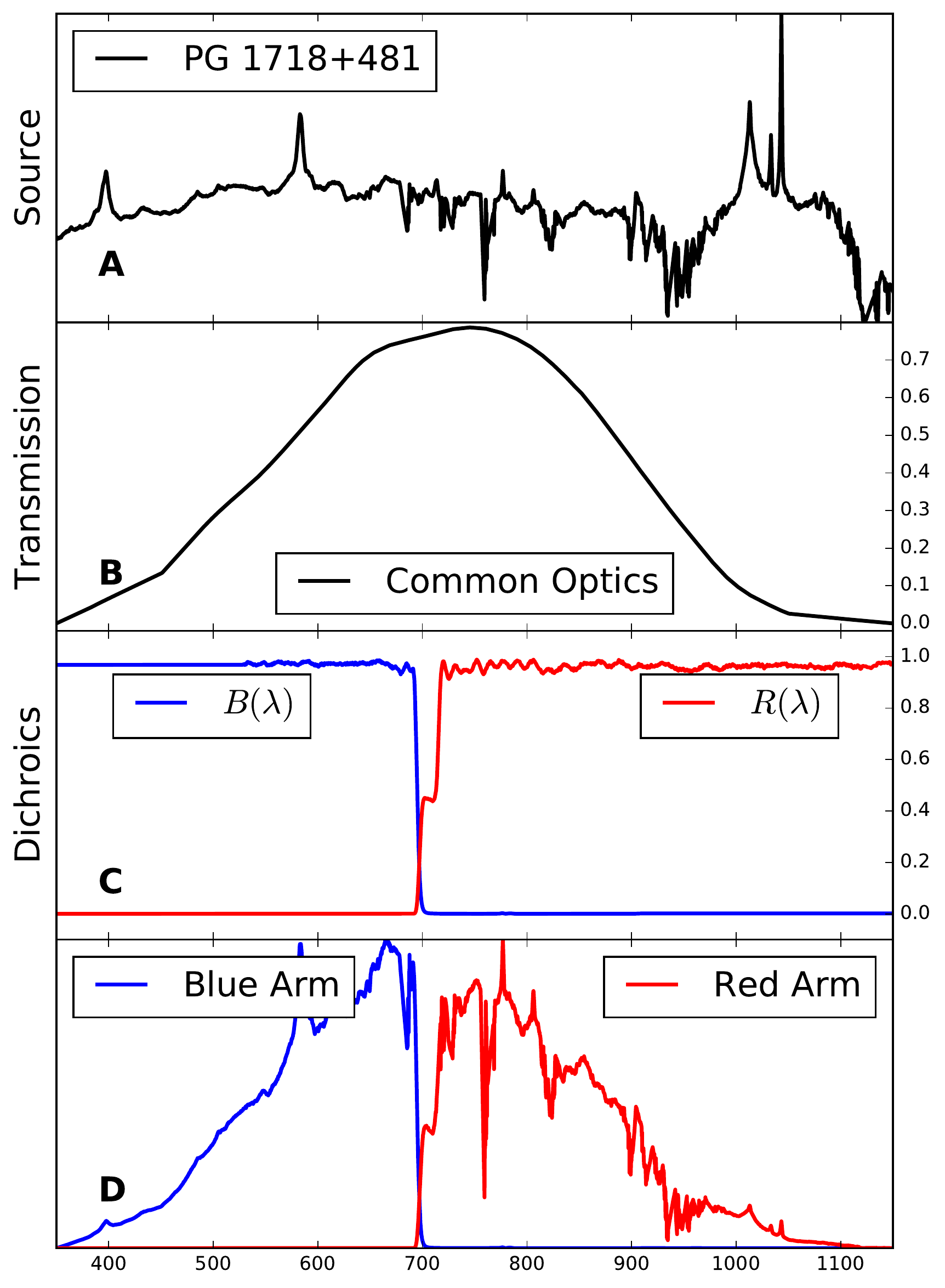}
    \caption{A: The atmosphere-attenuated spectrum of a typical quasar. B: The cumulative transmission curves of two lenses and the detectors. C: The splitting of photons down the blue/red arms induced by the dichroic beamsplitters. D: The product of curves in A-C gives the number distribution of photon colors at each arm, from which $f_{j \to j'}$ can be computed. 
    }
    \label{fig:fourPanelPlot}
\end{figure}

The parameter $f_{j\to j'}$ depends on the choice of what cutoff wavelength $\lambda'$ we choose to distinguish the photons we call red ($\lambda > \lambda'$) from blue ($\lambda < \lambda'$). It also depends on the emission spectra of the astronomical source. Note that since this color cutoff is completely arbitrary, we may choose $\lambda'$ differently for each astronomical source such that the crosstalk probability is minimized. These probabilities can be computed from the atmospheric scattering and absorption, detector quantum efficiencies, and transmission/reflection probabilities of the optics in each detector arm (see Fig.~\ref{fig:fourPanelPlot}). We define the following quantities, which all are dependent on wavelength:

\begin{indenteddesc}
 \item[$N_{\rm source} (\lambda)$] Number distribution of astronomical photons per wavelength that impinge on the top of Earth's atmosphere towards the telescope. We treat the interstellar/intergalactic medium as transparent because the column density of the ISM/IGM is lower than the Earth's atmosphere by at least a factor of 400, even over cosmological path lengths
 \item[$N_{\rm in}(\lambda)$] Number of photons per wavelength that are transmitted through the atmosphere and impinge
 on the pinhole mirror.
 \item[$\rho_{\rm lens}(\lambda)$] Probability of transmission through the collimating or focusing lens.
 \item[$\rho_{\rm det}(\lambda)$] Probability of detection by the APD (quantum efficiency).
 \item[$R(\lambda),B(\lambda)$] Probability of entering the red/blue arm due to the dichroic beamsplitters.
\end{indenteddesc}
In terms of these quantities, we can compute the overall spectral response of the red/blue arms of the instrument:
\begin{equation}
\begin{split}
    \rho_{\blue} &= B \times \rho_{\rm lens}^2 \times \rho_{\rm det} \nonumber \\
    \rho_{\red}  &= R \times \rho_{\rm lens}^2 \times \rho_{\rm det} \nonumber
\end{split}
\end{equation}
as well as the parameters $f_{j\to j'}$:
\begin{equation}
f_{b \to r} = \dfrac{\int_{0}^{\lambda'} N_{\rm in} R ~d\lambda}{\int_{0}^{\infty} N_{\rm in} R ~d\lambda} \, , \quad
f_{r \to b} = \dfrac{\int_{\lambda'}^{\infty} N_{\rm in} B ~d\lambda}{\int_{0}^{\infty} N_{\rm in} B ~d\lambda}\, .
\end{equation}

For bright stars such as the ones we observe, the quantity $N_{\rm source}(\lambda)$ is well-approximated by a blackbody \cite{ballesteros2012new}. For dim, redshifted quasars, we apply the appropriate Doppler shift to the composite rest-frame spectrum computed in Ref.~\cite{berk2001composite}. Once $N_{\rm source}$ is obtained, we compute $N_{\rm in}(\lambda)$ via the equation
\begin{equation} 
N_{\rm in}/N_{\rm source} = \rho_{\rm atm}(\lambda)\exp(-X \tau(\lambda))
\label{airmass}
\end{equation}
where $\rho_{\rm atm}(\lambda)$ is taken from the atmospheric radiative transfer code MODTRAN~\cite{berk2014modtran6} and takes into account the Rayleigh scattering and atmospheric absorption at zenith. In order to correct for off-zenith observations, we insert a factor of $\exp(-X \tau(\lambda))$ where $X$ is the observation airmass and $\tau(\lambda)$ is the optical depth due to Rayleigh scattering. In doing so, we make the approximation that the contribution to $f_{j \to j'}$ due to the optical density of absorption is negligible compared to Rayleigh scattering. 

In preparing for the recent cosmic Bell experiment~\cite{handsteiner2017cosmic}, it was determined that two dichroics were necessary because a single dichroic's optical density was low enough such that a non-negligible fraction of the light could go either way and would not be determined by the astronomical object. With this model, we selected our two dichroic beamsplitters to minimize the total amount of crosstalk while splitting the detector's sensitivity band in roughly equal halves. We determined that putting the short-pass dichroic beamsplitter first yielded lower crosstalk than the other way around. We used a 697\,nm short-pass dichroic beamsplitter (Semrock F697-SDi01-25x36) and an additional 705\,nm long-pass dichroic beamsplitter (Semrock FF705-Di01-25x36) to reduce the number of wrong-way photons in the reflected (red) arm.

For the quasars listed in Table \ref{tab:quasardata}, we compute $f_{j\to j'}$ values in the ranges $0.16\% < f_{b\to r} < 0.20\%$ and $0.17\% < f_{r\to b} < 0.23\%$, an order of magnitude better than the values of $f_{j \to j'}$ achieved with the instrumentation used for the original cosmic Bell experiment in Ref.~\cite{handsteiner2017cosmic}. We plot in Fig.~\ref{fig:fourPanelPlot}D the products $\rho_{\text{blue}} N_{\rm in}$ and $\rho_{\text{red}} N_{\rm in}$, where $N_{\rm in}$ is computed for the quasar PG 1718+481 at an observation altitude of 67 degrees.

\section{Observations}
\label{sec:calibration}

\begin{figure}
    \centering
    \includegraphics[width = 3.375in]{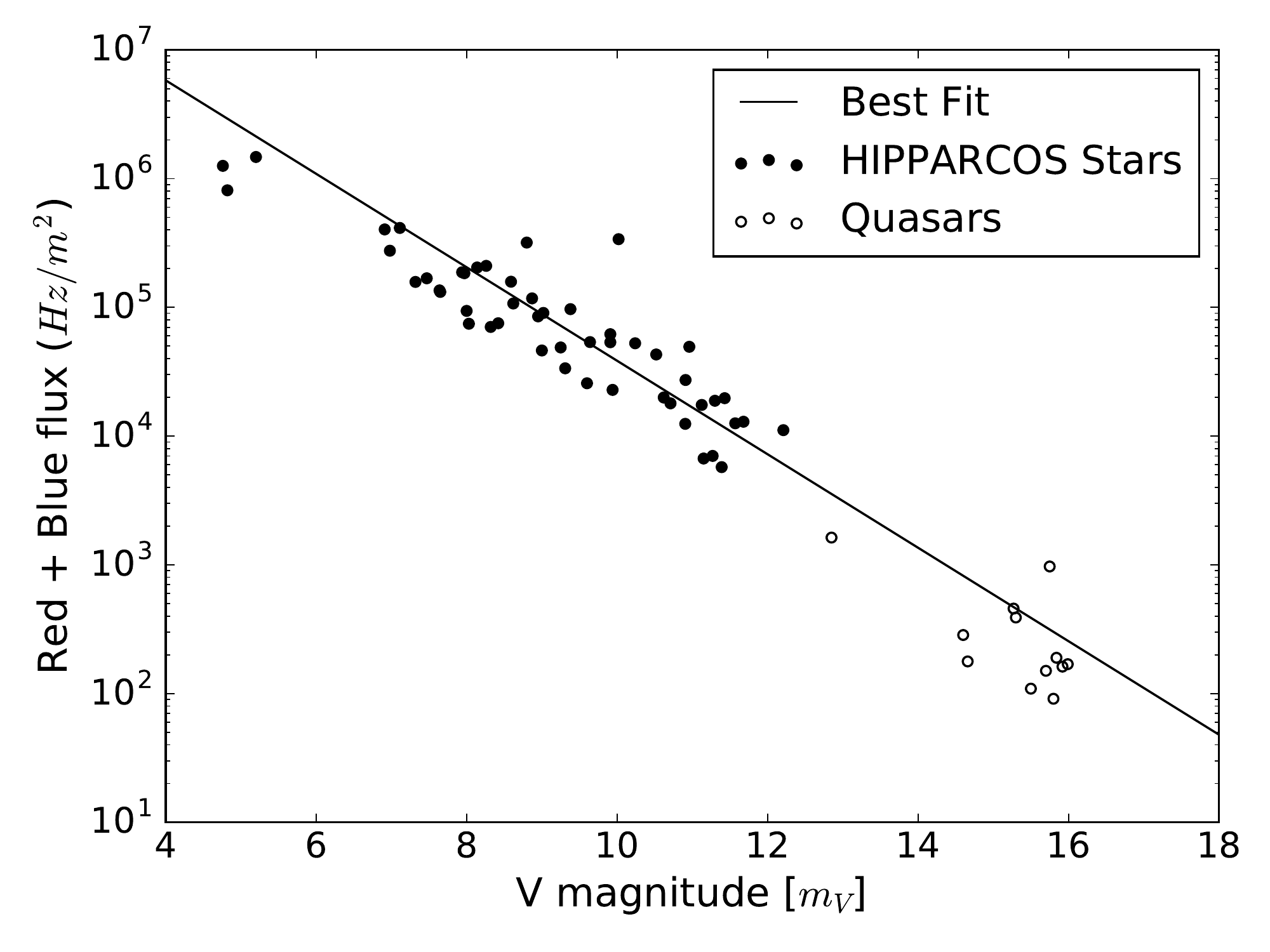}
    \caption{
For 50 bright stars in the HIPPARCOS catalog observed at zenith and 
twelve high-redshift quasars ($z < 3.911$), we plot the total (red + blue) background-subtracted count rate against the V-band magnitude (551$\pm$88\,nm). Though the V-magnitude is well into our blue band, it is the only data available for all observed objects and turns out to be a good predictor of the observed photon flux, as seen by the best-fit line that relates the two. We see subtle evidence of detector nonlinearity at high count rates, as discussed in the text.}
    \label{fig:magnitudeCalibration}
\end{figure}

We observed roughly 50 stars of varying B-V color roughly at zenith, generating astronomical random bits at rates from $\sim \SI{1e4}{\hertz} - \SI{1e6}{\hertz}$. Count rates for these, along with 12 different quasars, are plotted in Fig.~\ref{fig:magnitudeCalibration} as a function of astronomical V-band magnitude, denoted $m_V$. The V-band is defined by a broad filter centered at 551\,nm with a FWHM of 88\,nm.

Count rates as a function of time for dark counts and several stars and quasars are shown in Fig.~\ref{fig:countsOverTime}. To characterize the dark-count rates of the instrument, we close the telescope dome and obstruct its aperture with a tarp, and measure the count rate for about 500 seconds. We find that the variability in count rates, when integrated over 1 second, is consistent with a Poisson process with variance $\sqrt{N}$: in the blue arm we see $41$ cps, and in the red arm we see $93$ cps. At zenith, the background rates due to skyglow were roughly $\SI{20}{\hertz}$ and $\SI{60}{\hertz}$ in the blue and red arms respectively. (For comparison, the quasars we observed had rates of 100 to 1000\,Hz in each channel.) The reason for this asymmetry results from a combination of different optical coupling efficiencies in each arm and the spectrum of the background skyglow, which tends to be brighter in the near-infrared than in the visible band.

A comprehensive list of our star observations is available upon request. We find that the astronomical bit rate per telescope area is given approximately by 
\begin{align}
\nonumber &\log_{10}(\textrm{count rate [Hz / $m^2$]}) \\
&\quad\quad = (8.22 \pm 0.02) - (0.3631 \pm 0.0002) m_V
\end{align}
after subtracting skyglow and dark counts. The deviation from the expected slope of $-0.4$ is likely due to detectors becoming significantly saturated at count rates higher than $\sim \SI{1e5}{\hertz}$. 

In addition, we generated random bits from a number of quasars, with V band magnitudes ranging from 12.9 to 16, and redshifts up to $z = 3.911$, with bitrates ranging from $\sim \SI{1e2}{\hertz}-\SI{2e3}{\hertz}$. Light travel times $\tau$ are calculated from the maximally-constrained cosmological parameters from the Planck satellite \cite{ade2016planck}. The two most distant quasars we observed emitted their light over 12 billion years ago, a significant fraction of the 13.8 billion-year age of the universe. A summary of our quasar observations, and two measures quantifying the physical and information-theoretic predictability of bits ($p^{(i)}$ and $I$), are presented in Table \ref{tab:quasardata}. Timestamped random bits generated from these quasars are available at \url{https://stuff.mit.edu/~calvinl/quasar-bits/}.

\begin{table*}[t]
\centering
\begin{tabular}{m{0.16\textwidth}|m{0.10\textwidth}|m{0.08\textwidth}|m{0.08\textwidth}|m{0.07\textwidth}|m{0.08\textwidth}|m{0.08\textwidth}|m{0.14\textwidth}|m{0.13\textwidth}}
Name        & Redshift (z) & $\tau$ (Gyr) & B & V & blue (cps) & red (cps) & valid fraction $q^{(i)}$ & max info $I \times 10^4$\\
\hline
3C 273             & 0.173 & 2.219  & 13.05  & 12.85  & 672    & 1900    & 0.884 & 87.8\\ 
HS 2154+2228       & 1.29  & 8.963  & 15.2   & 15.30  & 227    & 503     & 0.774 & 9.91\\
MARK 813           & 0.111 & 1.484  & 15.42  & 15.27  & 193    & 633     & 0.703 & 7.62\\
PG 1718+481        & 1.083 & 8.271  & 15.33  & 14.6   & 176    & 473     & 0.682 & 3.07\\
APM 08279+5255     & 3.911 & 12.225 & 19.2   & 15.2   & 684    & 1070    & 0.647 & 5.39\\ 
PG1634+706         & 1.337 & 9.101  & 14.9   & 14.66  & 121    & 285     & 0.572 & 3.38\\
B1422+231          & 3.62  & 12.074 & 16.77  & 15.84  & 123    & 358     & 0.507 & 4.22\\
HS 1603+3820       & 2.54  & 11.234 & 16.37  & 15.99  & 121    & 326     & 0.501 & 4.78\\
J1521+5202         & 2.208 & 10.833 & 16.02  & 15.7   & 106    & 309     & 0.476 & 2.39 \\
87 GB 19483+5033   & 1.929 & 10.409 & unknown& 15.5   & 98     & 241     & 0.464 & 0.32\\
PG 1247+268        & 2.048 & 10.601 & 16.12  & 15.92  & 111    & 333     & 0.453 & 2.92\\
HS 1626+6433       & 2.32  & 10.979 & unknown& 15.8   & 87     & 213     & 0.398 & 1.81\\
\end{tabular}
\caption{A list of quasars observed, their corresponding redshifts $z$, and light travel times $\tau$. We report their B and V magnitudes from the SIMBAD Astronomical Database and our observed $75$th percentile count rates. The table is sorted by the fraction of valid settings $q^{(i)}$ for each quasar observation, based on both off-target counts measured at each observation's airmass and rates for quasar photons to go the wrong way through our imperfect dichroics calculated from each quasar's emission spectrum. Predictability, as measured by $I = \max_m I_N(m;m+1)$, is the small mutual information we measured in each quasar's bitstream and corresponds to a negligible reduction in $q^{(i)}$. Even using a small ($\SI{1}{\m}$) telescope at a light-polluted Los Angeles observing site, we find that the first quasar (3C 273) paired with either of the next two would yield $q^{\rm Alice} + q^{\rm Bob}$ in excess of the limit set by Eq.~(\ref{qAqB}) for addressing the freedom-of-choice loophole.
}
\label{tab:quasardata}
\end{table*}

\section{Quality of Randomness}
\label{sec:quality}

In addition to quantifying the fraction of valid runs as was done in Ref.~\cite{handsteiner2017cosmic}, we may assess the quality of randomness statistically to yield a measure of predictability. The NIST Statistical Test Suite \cite{rukhin2001statistical} provides a device-independent statistical approach to evaluate the quality of the output of any random number generator given a sufficiently large number of bits. When using timestamps to generate random bits based on whether photons arrive on an even or odd nanosecond, we find that our random numbers pass the NIST test suite, consistent with the findings in Ref.~\cite{Wu2016}. When using photon colors to generate random bits, our data fail the NIST tests, largely due to the existence of an overall imbalance in red-blue count rates. 

To quantify imperfect statistical randomness in a bitstream, we may consider the mutual information between a moving window of $m$ bits and the $(m + 1)$th bit, which we denote as $I(m;m+1)$. If each bit were truly independent, this mutual information would be zero, even if the probability to get a 0 or 1 was not 50\%. To define $I(m; m+1)$, let $\mathcal{X}_m$ denote the set of all length-$m$ binary strings, and let $p(x)$ be the probability that an $m$-bit string within our bitstream is $x\in \mathcal{X}_m$. Similarly, let $p(y)$ be the probability that the next bit is $y \in \{0,1\}$. If we define $p(x,y)$ to be the probability that a string of $m+1$ bits are $x$ followed by $y$, then the mutual information in our data is defined to be
\begin{equation} 
I(m;m+1) = \sum_{x\in \mathcal{X}_m} \sum_{y\in \{0,1\}} p(x,y) \times \log_2 \left( \dfrac{p(x,y)}{p(x)p(y)} \right).
\label{eq:mi_def}
\end{equation}
Note that if the next bit is independent of the $m$ bits preceding it, then $p(x,y)=p(x)p(y)$ and the mutual information vanishes. 

Estimating the true mutual information in a sample of length $N$, denoted $I_N(m;m+1)$, is in general a highly non-trivial problem. Precise knowledge of the true probabilities $p(x,y),p(x)$, and $p(y)$ is required. While statistical fluctuations in counting the numbers of zeros and ones in a particular dataset has an equal chance of overestimating or underestimating the finite-sample estimates $\hat{p}(x,y)$, $\hat{p}(x)$, and $\hat{p}(y)$, any statistical fluctuations in these probability estimates cause an upward bias in the estimated mutual information in the dataset~\cite{treves1995upward} if we simply ``plug in'' the experimental probability estimates $\hat{p}$ into Eq.~(\ref{eq:mi_def}), which takes as input the true probabilities $p$. An intuitive explanation for this bias in the mutual information is that our mutual information estimator cannot distinguish between a true pattern in the collected data and a random statistical fluctuation. We emphasize that it is statistical fluctuations in the count rates that cause overestimation of the mutual information. For example, a random realization of a 50/50 bitstream composed of 0's and 1's is unlikely to have exactly the same number of 0's and 1's (or the exact same number of occurrences of 01's and 00's), but regardless of whether there are more 01's or 00's, the mutual information will increase.

We denote this upward-biased estimator by $\hat{I}_N (m;m+1)$. However, in the limit that the dataset is large ($N \gg 1$), and if $m$ is fixed, the amount of positive bias in the estimated mutual information $\hat{I}_{N}(m;m+1)$ is dependent only on $N$ and can be represented as a perturbation away from the true mutual information $I(m;m+1)$. 
To construct an unbiased estimator that removes these finite-size effects, we adopt the ansatz~\cite{treves1995upward}
\begin{equation}
\hat{I}_{N}(m;m+1) = I(m;m+1) + \frac{a}{N} + \frac{b}{N^2} \, ,
\label{eq:mi_asymptotic}
\end{equation}
where $I(m;m+1)$, $a$, and $b$ are fixed, unknown constants, with finite-size effects being captured in values of $a$ and $b$. To determine these constants, we first compute $\hat{I}_{N}(m;m+1)$ for the entire dataset. By splitting the dataset into 2 chunks of size $N/2$, we may estimate $\hat{I}_{N/2}(m;m+1)$ by averaging the naive estimate from both chunks. Repeating this procedure for 4 chunks of size $N/4$ gives us a system of three equations linear in the unknowns $I(m;m+1)$, $a$, and $b$. 

\begin{figure}
    \centering
    \includegraphics[width = 3.375in]{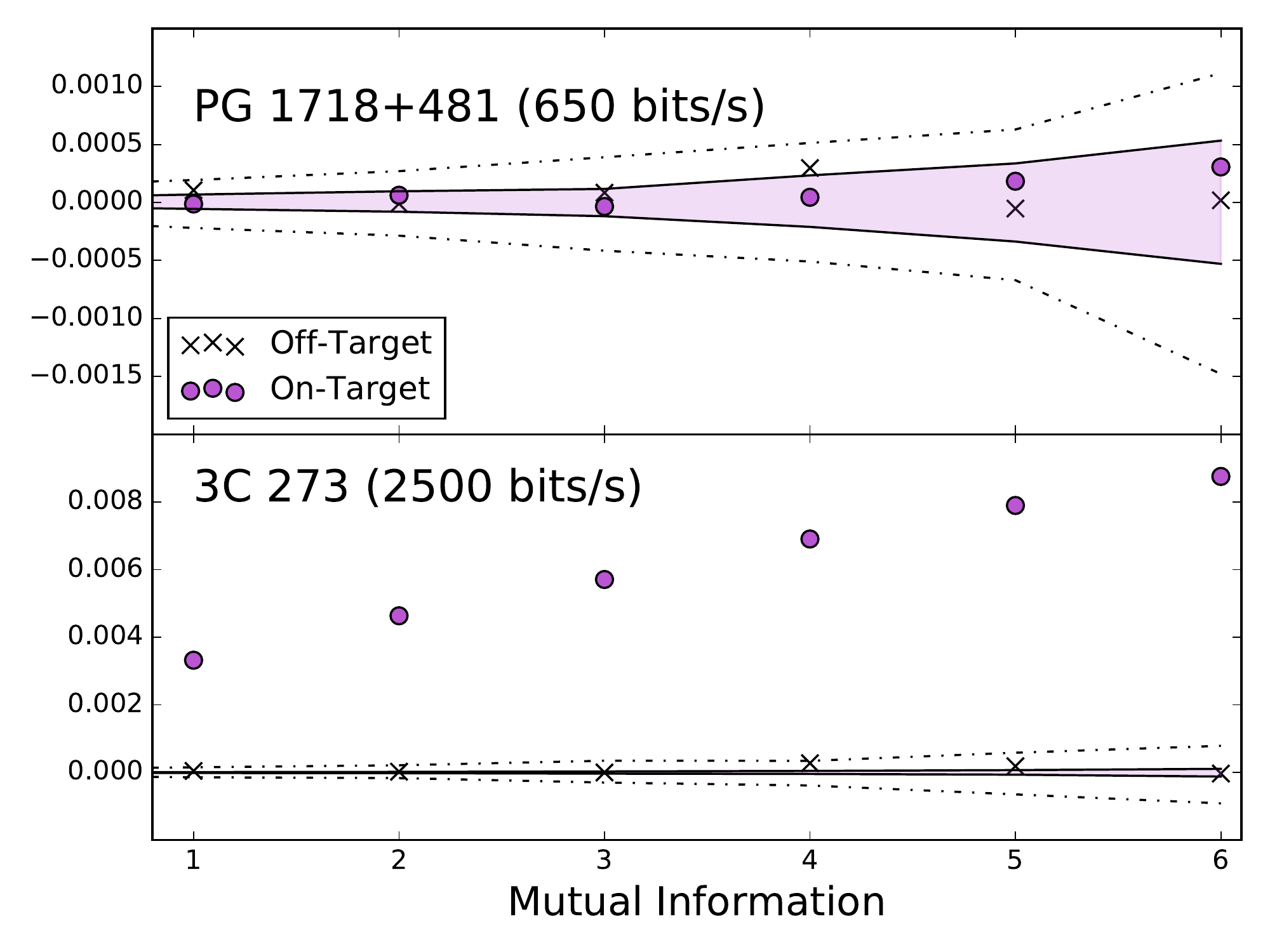}
    \caption{We plot for two different quasars our estimate of the mutual information between a bit and the $m$ bits preceding it, for $m = 1,\ldots,6$, for bitstreams generated when the quasar light is both present (circles) and absent (crosses). To test the hypothesis that the mutual information in our on-target data is consistent with zero, we analyze the mutual information of 50 pseudorandom bitstreams with the same length and red-blue bias as our astronomical bitstreams. The statistical distributions of the mutual information in the simulated bitstreams are shown for simulated on-target data (shaded, purple fill) and simulated off-target noise (dotted lines). For the quasar PG1718+481, we find that the experimentally observed mutual information in the on-target as well as the off-target data is consistent with zero, while the mutual-information deviates significantly from zero when observing the exceptionally bright quasar 3C 273 ($\approx 2500$cps). In both cases, data taken off target never exceeds 200 cps. This illustrates how increased bitrates lead to slightly increased statistical predictability, as discussed in the text.}
    \label{fig:mi_good}
\end{figure}

From this procedure, we compute an unbiased estimate of the mutual information in the bits we generate when taking on-quasar data as well as data taken when pointing at the sky slightly off-target. We compute $I(m;m+1)$ for $m = 1,2,\ldots 6$ lookback bits on our datasets of sufficient length $N > 2^{16}$ to run. To determine whether our estimates are consistent with zero mutual information, we compare our estimates of $I_{N}(m;m+1)$ against fifty simulated datasets, each with the same length and the same red-blue imbalance but with no mutual information. Examples of a quasar bitstream with almost no mutual information (PG 1718+481) and a quasar bitstream with nonzero mutual information (3C 273) are shown in Fig.~\ref{fig:mi_good}.

For the quasars in Table~\ref{tab:quasardata}, we observe that the random bits generated from colors in 8 out of 12 datasets exhibit mutual information that is statistically significantly different from zero, though still very small. This hints at the possibility of some nontrivial structure in the data which may be induced by physical effects or systematic error. For the exceptionally bright quasar 3C 273 ($V = 12.9$), we measure $I \approx 0.009$, while in the remaining 11 datasets, the maximum mutual information $I = \max_m I(m;m+1)$ never exceeds $0.001$. One way to realize a mutual information of $0.001$ is to have one in every 1000 bits be a deterministic function of the previous few bits instead of being random. Even in the worst case of $0.009$, the amount of predictability is only increased negligibly compared to the effect from skyglow, and is well below the threshold needed to address the freedom-of-choice loophole in a Bell test.  For example, in the recent cosmic Bell experiment \cite{handsteiner2017cosmic}, violations of the Bell-CHSH inequality were found with high statistical significance ($>$7 standard deviations) for an experiment involving $\sim 10^5$ detected pairs of entangled photons, even with excess predictability in each arm of each detector of order $p^{(i)} \sim 0.1$.

Upon examining the experimental probability estimates $\hat{p}(x,y)$ that went into the mutual information calculation, we identified two systematic sources of non-randomness, both of which are exacerbated at high bitrates. The first mechanism for non-randomness is detector saturation. After a detection, the detector has a hard-coded $\SI{420}{\ns}$ deadtime window during which a detection is improbable. Hence for sufficiently high count rates (such as those experienced when observing stars), it is much more likely to observe a blue photon following a red one and vice versa than multiple photons of the same color in a row. While we see this effect in our calibration data with HIPPARCOS stars, the count rates necessary for this effect to be important ($10^5-10^6$ counts per second) far exceed what is observed with quasars. These are eliminated by imposing the same deadtime window in the other channel and removing (in real time or in post-processing) any detection that is within the deadtime of any previous detection from either channel.

The second mechanism is a consequence of imperfect alignment combined with random atmospheric seeing. The exact extent of a slight geometric misalignment is extremely difficult to measure and changes slightly day to day. We checked the optical alignment before each night of observation, and the device's alignment remained quite stable from night to night for over a week|a practical boon for a cosmic Bell test. However, due to our device's imperfectly-manufactured pinhole, we know there exists a ``sweet spot'' for optimal coupling to the blue detector, and a slightly different sweet spot for optimal alignment with the red detector. As the image of the quasar twinkles within the pinhole on timescales of milliseconds, its instantaneous scintillation pattern overlaps differently with these sweet spots. The result is that when photon fluxes increase to rates approaching one per millisecond ($\sim 1000$cps), the conditional probability $p(x\to y)$ of receiving detection $y$ given previous detections $x$ begins to exceed the average probability of obtaining $y$ if the last few bits in $x$ are the same as $y$. For example, for quasar 3C273 we see $p(10111\to 1) = p(101111) / p(10111) = 0.751 > p(1) = 0.726$. We suspect that this effect is responsible for the nonzero statistical predictability in our data and leads to an increased predictability of a few parts in $10^4$ for high count rates.

Since atmospheric seeing is a consequence of random atmospheric turbulence, it is a potential source of local influences on astronomical randomness. It can be mitigated by careful characterization of the optical alignment of the system, making sure that the sweet spots of both detector arms overlap to the greatest extent possible, using detectors with a large, identical active detector area, and observing under calm atmospheric conditions. For a larger telescope in a darker location where the signal to background ratio is higher, this would be a relatively larger effect on the fraction of valid runs.

\section{Conclusion}
\label{sec:discussion}

Building on the design and implementation of astronomical random number generators in the recent cosmic Bell experiment \cite{handsteiner2017cosmic}, we
have demonstrated the capabilities of a telescope instrument that can output a time-tagged bitstream of random bits based on the detection of single photons from astronomical sources with tens of nanoseconds of latency. We have further demonstrated its feasibility as a source of random settings for such applications as testing foundational questions in quantum mechanics, including asymptotically closing the freedom-of-choice loophole in tests of Bell's inequality, and conducting a cosmic-scale delayed-choice quantum-eraser experiment. Beyond such foundational tests, astronomical sources of random numbers could also be of significant use in quantum-cryptographic applications akin to those described in Refs.~\cite{Wu2016,colbeck12,barrett05,pironio09,pironio10,vazirani14,gallego13}.

Other interesting applications of this device may be found in high time-resolution astrophysics. For example, it might be possible to indirectly detect gravitational waves and thereby perform tests of general relativity with the careful observation of several optical pulsars using future versions of our instrument and larger telescopes, complementing approaches described in Refs.~\cite{hobbs13,mclaughlin13,kramer13,manchester13,lazio13}. 

\section*{Acknowledgements}

We are grateful to members of the cosmic Bell collaboration for sharing ideas and suggestions regarding the instrument and analyses discussed here, and for helpful comments on the manuscript, especially Johannes Handsteiner, Dominik Rauch, Thomas Scheidl, Bo Liu, and Anton Zeilinger.
Heath Rhoades and the other JPL staff at Table Mountain were invaluable for observations. We also acknowledge Michael J.~W. Hall for helpful discussions, and Beili Hu, Sophia Harris, and an anonymous referee for providing valuable feedback on the manuscript. Funding for hardware and support for CL and AB was provided by JG's Harvey Mudd startup. This research was carried out partly at the Jet Propulsion Laboratory, California Institute of Technology, under a contract with the National Aeronautics and Space Administration and funded through the internal Research and Technology Development program. This work was also supported in part by NSF INSPIRE Grant No. PHY-1541160. Portions of this work were conducted in MIT's Center for Theoretical Physics and supported in part by the U.S. Department of Energy under Contract No. DE-SC0012567.


\begin{thebibliography}{65}
\bibitem [{end()}]{endnotetext}%
\BibitemOpen
\href@noop {} {}\bibinfo {note} {{Single-photon delayed-choice experiments
		focus on wave-particle duality. However, a determined skeptic of quantum
		theory could concoct a local-realist explanation for the outcomes. Such an
		explanation would require that two local-hidden-variable-like surrogates of
		the photon that enters the interferometer each travel around one path,
		accumulating a phase based on their distance traveled. When they come
		together, they would either see a beamsplitter or not. They could either
		combine their accumulated phases and act like a wave or they could ignore
		their phases and pick one detector over the other in some deterministic or
		locally-probabilistic way. In this way, there would exist a perfectly
		local-realist explanation for the wave-particle duality manifested by single
		particles. On the other hand, the outcomes of two-photon experiments such as
		a delayed-choice quantum eraser cannot be accounted for within a
		local-realist framework}}\BibitemShut {NoStop}%
\end{thebibliography}



%

\end{document}